\documentclass[10pt]{article}
\usepackage{amsmath,amsfonts,epsfig,amssymb}

\newcommand{\er}{\mathbb{R}}

\newcommand{\dx}{\mbox{d}x}

\newcommand{\dxi}{\mbox{d}\xi}

\newcommand{\du}{\mbox{d}u}
\newcommand{\cros}{{\cal X}}
\newcommand{\squ}{{\cal S}}

\newcommand{\setP}{{\cal P}}
\newcommand{\setQ}{{\cal Q}}
\newcommand{\setG}{{\cal G}}
\newcommand{\setH}{{\cal H}}

\newcommand{\dt}{\mbox{d}t}

\newcommand{\dd}{\mbox{d}}
\newcommand{\boldA}{{\mathbf A}}
\newcommand{\boldB}{{\mathbf B}}
\newcommand{\boldC}{{\mathbf C}}
\newcommand{\boldN}{{\mathbf N}}
\newcommand{\boldR}{{\mathbf R}}
\newcommand{\boldW}{{\mathbf W}}
\newcommand{\picturesAB}[3]{
\centerline{
\hspace*{.15in}
\raise 2mm \hbox{\raise #3 \hbox{(a)}}
\hspace*{-.3in}
\psfig{file=#1,height=#3}
\raise 2mm \hbox{\raise #3 \hbox{(b)}}
\hspace*{-.3in}
\psfig{file=#2,height=#3}
}}
\newcommand{\picturesABx}[6]{
\centerline{\raise #3 \hbox{#5}
\psfig{file=#1,height=#3}
\hspace*{#4}
\raise #3 \hbox{#6}
\psfig{file=#2,height=#3}
}}
\newcommand{\picturesABm}[5]{
\centerline{\raise #3 \hbox{(a)}
\hspace*{#5}
\psfig{file=#1,height=#3}
\hspace*{#4}
\raise #3 \hbox{(b)}
\hspace*{#5}
\psfig{file=#2,height=#3}
}}
\newcommand{\picturesABCm}[4]{
\centerline{\raise #4 \hbox{(a)}\!\!\!\!\!
\psfig{file=#1,height=#4}
\hspace*{.06in}
\raise #4 \hbox{(b)}\!\!\!\!\!\!\!\!\!\!\!\!
\psfig{file=#2,height=#4}
\hspace*{.06in}
\raise #4\hbox{(c)}\!\!\!\!\!\!\!\!\!\!\!\!
\psfig{file=#3,height=#4}
}}
\newcommand{\picturesABCD}[5]{
\centerline{\raise #5 \hbox{(a)}\!\!\!
\psfig{file=#1,height=#5}
\hspace*{.2in}
\raise #5 \hbox{(b)}\!\!\!
\psfig{file=#2,height=#5}
\hspace*{.2in}
\raise #5\hbox{(c)}\!\!\!
\psfig{file=#3,height=#5}
\hspace*{.2in}
\raise #5\hbox{(d)}\!\!\!
\psfig{file=#4,height=#5}
}}

\numberwithin{equation}{section}

\begin{document}

\title{On chemisorption of polymers to solid surfaces}

\author{
Radek Erban\!\,\thanks{University of Oxford, Mathematical Institute, 
24-29 St. Giles', Oxford, OX1 3LB, United Kingdom;
e-mail: {\it erban@maths.ox.ac.uk}. This work was supported 
by Biotechnology and Biological 
Sciences Research Council.}
 \and 
Jonathan Chapman\!\,\thanks{University of Oxford, Mathematical Institute, 
24-29 St. Giles', Oxford, OX1 3LB, United Kingdom;
e-mail: {\it chapman@maths.ox.ac.uk}.}
}

\date{\small \today}

\maketitle

\leftskip 1cm
\rightskip 1cm

\noindent {\bf Abstract.}
The irreversible adsorption of polymers to a two-dimen\-sional solid 
surface is studied.  An operator formalism is introduced for 
chemi\-sorp\-tion from a polydisperse solution of polymers which transforms 
the analysis of the adsorption process to a set of combinatorial problems 
on a two-dimensional lattice. The time evolution of the number of polymers 
attached and the surface area covered are calculated via a series
expansion. The dependence of the final
coverage on the parameters of the model (i.e. the parameters of the
distribution of polymer lengths in the solution) is studied. 
Various methods for accelerating the convergence of the resulting 
infinite series are considered. 
To demonstrate the accuracy of the truncated series approach, 
the series expansion results are compared with the results of stochastic 
simulation.
 
\leftskip 0cm
\rightskip 0cm

\section{Introduction} 

\label{secintro}

\begin{figure}
\centerline{
\raise 0.92in \hbox{(a)}
\hskip -4mm
\psfig{file=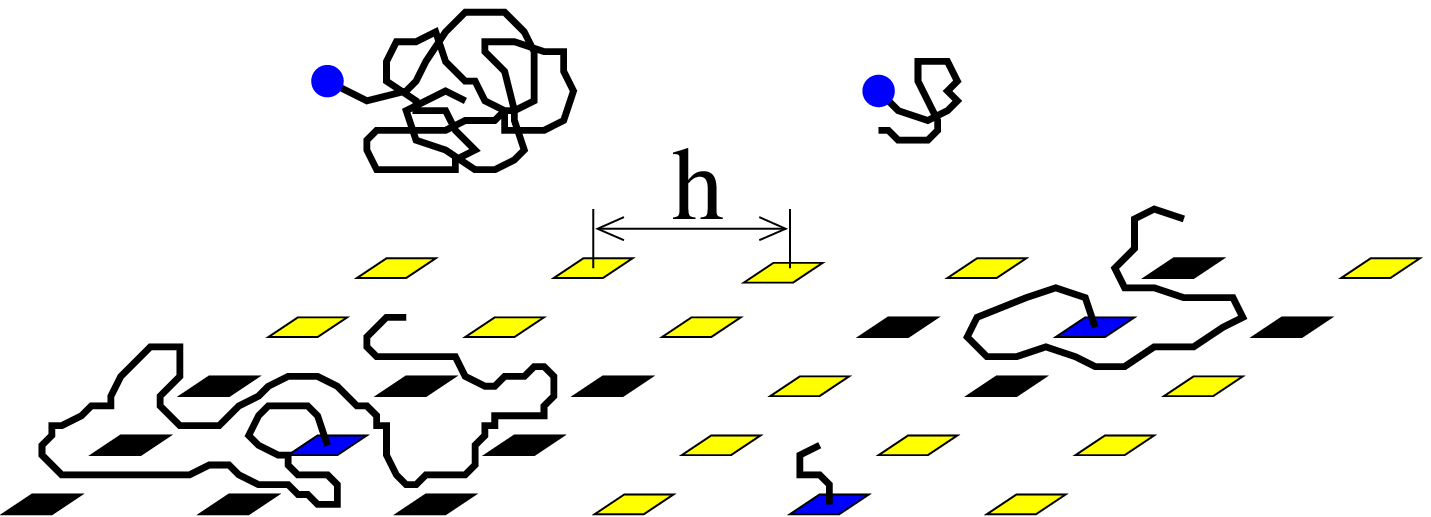,height=1.1 in,width=3.3in}
\hskip 8mm
\raise 0.92in \hbox{(b)}
\psfig{file=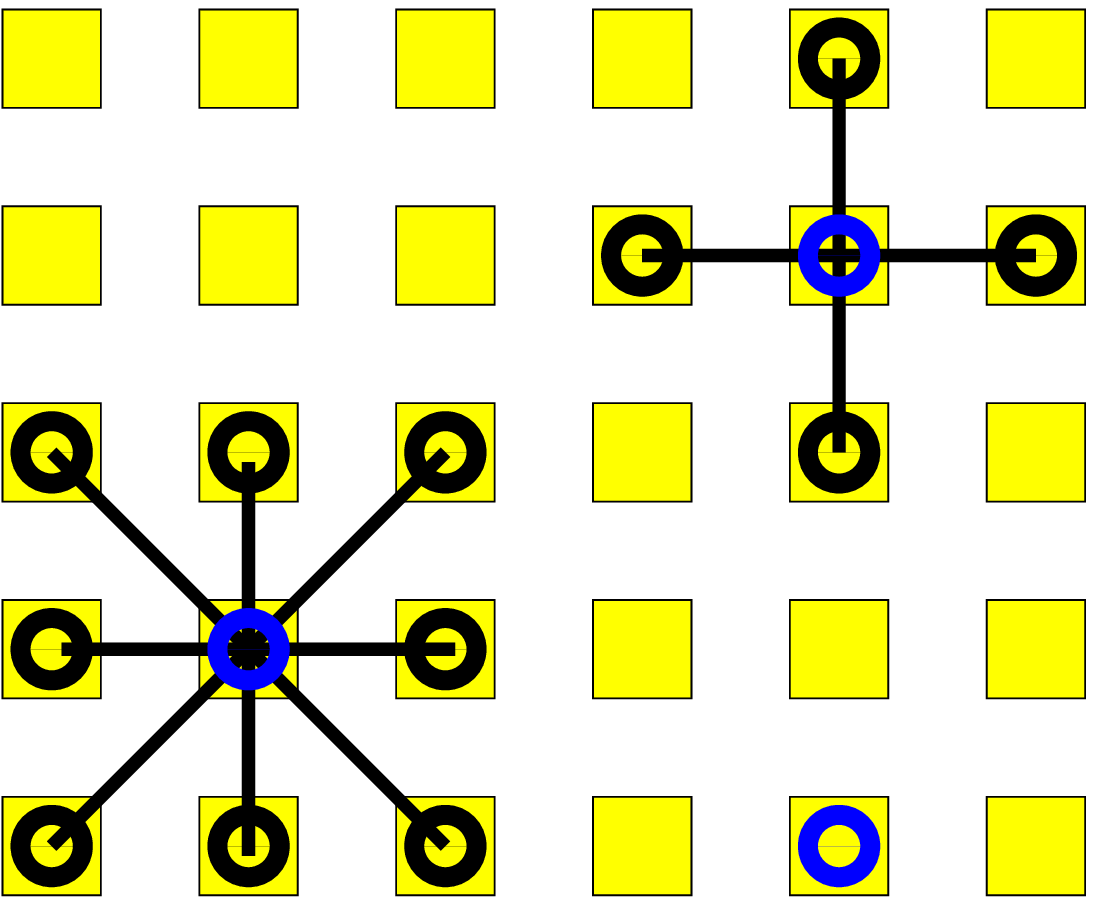,height=0.8 in}
}
\caption{(a) {\it A schematic diagram of the chemisorption of the 
semitelechelic polymer.}
(b) {\it Situation from (a) transformed into two-dimensional
lattice setting.}}
\label{figschematic1}
\end{figure}

The adsorption of polymers to solid surfaces has wide 
technological and medical applications  
\cite{Shaughnessy:2005:NAP,Fisher:2001:PCA}. In this paper,
we study chemisorption, i.e. the situation where covalent 
surface-polymer bonds develop and adsorption is effectively
irreversible on the experimental time scale \cite{Shaughnessy:2003:IPA}.
Chemisorbing polymers have one or more reactive (binding) groups
along the polymer chain which can react with binding sites on the surface.
Polymers with one reactive group at the end of the chain are
called semitelechelic. A schematic diagram of the adsorption of a 
semitelechelic polymer is shown in Figure \ref{figschematic1}(a) where 
the binding sites are arranged into a rectangular mesh on the surface. 
An important 
parameter of the chemisorption process is the density of binding sites, or 
equivalently, the average distance between neigbouring sites, which
is denoted by $h$ in Figure \ref{figschematic1}(a).
Denoting the hydrodynamic radius of the polymer by $R$, we can distinguish
three different scenarios. If $h \ll R$, then the polymer layer created 
by chemisorption of the semitelechelic polymer will be
a polymer brush after sufficiently long time 
\cite{Milner:1988:TGP,Himmelhaus:2003:GDP,Zajac:1994:KTE}.
In this case, one can simply assume that a polymer can attach anywhere 
on the surface for modelling purposes. In particular, one can use 
continuum random sequential adsorption to model 
the process \cite{Erban:2006:DPI}. The other extreme case
is $h \gg R$ where the final layer contains one attached polymer  
at each binding site. No steric shielding needs to be considered
when modelling the process and the dynamics of adsorption is trivial 
from the mathematical
point of view. The last important case is when $h \sim R$. This is the
regime studied in this paper.

Chemisorption is often modelled as a random sequential 
adsorption (RSA) \cite{Evans:1993:RCS,Talbot:2000:CPP}. In a previous 
paper \cite{Erban:2006:DPI} 
we studied one-dimensional models of random sequential (irreversible) 
adsorption. Our motivation was to understand the essential processes involved 
in pharmacological applications such as the polymer coating of viruses
\cite{Fisher:2001:PCA}. The classical RSA model \cite{Evans:1993:RCS} 
was generalized to study the effects 
of polydispersity of polymers in solution, of partial overlapping of the 
adsorbed polymers, and the influence of reactions with the solvent
on the adsorption process. Working in one dimension, we derived
an integro-differential evolution equation for the adsorption
process and we studied the asymptotic behaviour of the quantities 
of interest, namely the surface area covered and the number of molecules 
attached to the surface. We also presented applications of equation-free 
dynamics renormalization tools \cite{Kevrekidis:2003:EFM} to study the 
asymptotically self-similar behaviour of the adsorption process.
In \cite{Erban:2006:DPI} we used a continuum RSA model. 
The underlying assumption was that the polymer
can effectively bind anywhere on the surface, i.e. we worked
in the regime $h \ll R.$ In reality, the reactive groups on the polymer
can react only with the corresponding binding sites on the surface,
which are primary amino-groups in the virus coating problem.
Rough estimates from molecular models suggest that the average distance 
between primary amino-groups in the virus capsid is about
a nanometre \cite{Fisher:2005:PC}. However, it is difficult to guess 
which of the amino-groups in the capsid are available for the reaction 
with the polymer, i.e. are accessible for polymers from solution. 
In particular, both the regimes $h \ll R$ and $h \sim R$ can be 
justified in the virus coating problem. Other chemisorbing systems
\cite{Shaughnessy:2005:NAP,Evans:1993:RCS} can be also used
to motivate investigation of the borderline case $h \sim R$.

Assuming $h \sim R$, we have to take the discrete nature of the binding 
sites into 
account. This means that lattice RSA modelling is more appropriate than 
continuum RSA modelling.
In this paper we assume for simplicity that the binding sites lie
on a rectangular mesh (see Figure \ref{figschematic1}),
with mesh points a distance $h$ apart. We choose $h=1$ without loss
of generality in what follows. Any polymer covers 
the binding site to which it is attached. Moreover, longer polymers also 
effectively cover neighbouring binding sites, as illustrated in Figure 
\ref{figschematic1}(a). More precisely, an attached semitelechelic 
polymer covers a circle of a certain radius $r$ which is centered at 
the binding site (meshpoint $(i,j)$). If $r < 1$, then the polymer effectively 
covers only the corresponding binding site $(i,j)$. 
If $1 \le r < \sqrt{2}$, then the polymer covers  a small ``cross" 
$\cros_{i,j}$ where we define 
\begin{equation}
\cros_{i,j} = \big\{ (i,j), (i+1,j), (i-1,j), (i,j-1), (i,j+1) \big\}.
\label{crossdef}
\end{equation}
We call set of mesh points $\cros_{i,j}$ the cross (or cross-polymer) 
centered at $(i,j)$. If $\sqrt{2} \le r < 2$, then the polymer covers 
a small ``square" $\squ_{i,j}$ defined by
\begin{eqnarray}
\squ_{i,j} & = & \big\{ (i,j), (i+1,j), (i-1,j), (i,j-1), (i,j+1), 
\nonumber
\\
& & \;\; (i+1,j+1), (i+1,j-1), (i-1,j+1), (i-1,j-1) \big\}.
\label{squaredef}
\end{eqnarray}
We call set of mesh points $\squ_{i,j}$ the square (or square-polymer) 
centered at $(i,j)$. If $2 < r$, then the polymer covers 
at least 13 binding sites. To simplify the combinatorial complexity of 
the problem, we restrict our consideration to the case $r  < 2$. 
In this case, we can formulate the chemisorption 
of polymers in terms of adsorption of points, crosses and squares to 
the two-dimensional lattice (see Figure \ref{figschematic1}(b)). We denote  by
$\alpha$ the fraction of polymers in the solution for which 
$1 \le r < \sqrt{2}$,
so that $\alpha$ is the probability that a randomly chosen polymer 
in solution will adsorb as a cross. Similarly, we denote $\beta$ 
the fraction of polymers in the solution for which $\sqrt{2} \le r < 2$
so that $\beta$ is the probability that a randomly chosen polymer 
in solution will adsorb as a square. In particular, we must have
$0 \le \alpha + \beta \le 1$ where $1 - \alpha - \beta$ is the probability
that a randomly chosen polymer in solution will adsorb as a point.
We work with an $M \times M$ mesh with periodic boundary conditions. 
Then our two-dimensional polydisperse random sequential adsorption 
(pRSA) algorithm can be stated as follows.

\medskip

\leftskip 8mm
\rightskip 8mm

\noindent
{\bf pRSA algorithm:} 
{\it 
We consider the adsorption of points $\{(i,j)\}$, crosses $\cros_{i,j}$
and squares $\squ_{i,j}$ to the two-dimensional rectangular $M \times M$ mesh. 
At each time step, we choose randomly a point $(i,j)$ in the mesh. 
If the selected mesh point $(i,j)$ is covered (occupied) by 
a point/cross/square already placed, the adsorption is rejected. 
If the mesh point $(i,j)$ is vacant, then it is marked 
as occupied. 
Moreover, with probability $\alpha$ (resp. $\beta$), all mesh points 
in the set  $\cros_{i,j}$ (resp. $\squ_{i,j}$) are marked as occupied.
}

\leftskip 0mm
\rightskip 0mm

\medskip

\noindent
To simulate pRSA algorithm, we have to generate three random numbers
at each time step. The first two of them are used for random selection 
of the lattice point where the reactive group of the adsorbed polymer is 
attempted to bind. The third random number $r_n$, uniformly distributed
in interval $[0,1]$, is used to determine the length of the adsorbed
polymer. If $r_n \in [0,\alpha)$, then the cross polymer is placed.
If $r_n \in [\alpha, \alpha + \beta)$, then the square-polymer is
chosen. If $r_n \in [\alpha + \beta,1]$, then the point-polymer
is adsorbed. An illustrative numerical simulation of pRSA algorithm 
for $\alpha=0.8$ and $\beta=0.1$
\begin{figure}[ht]
\centerline{
\psfig{file=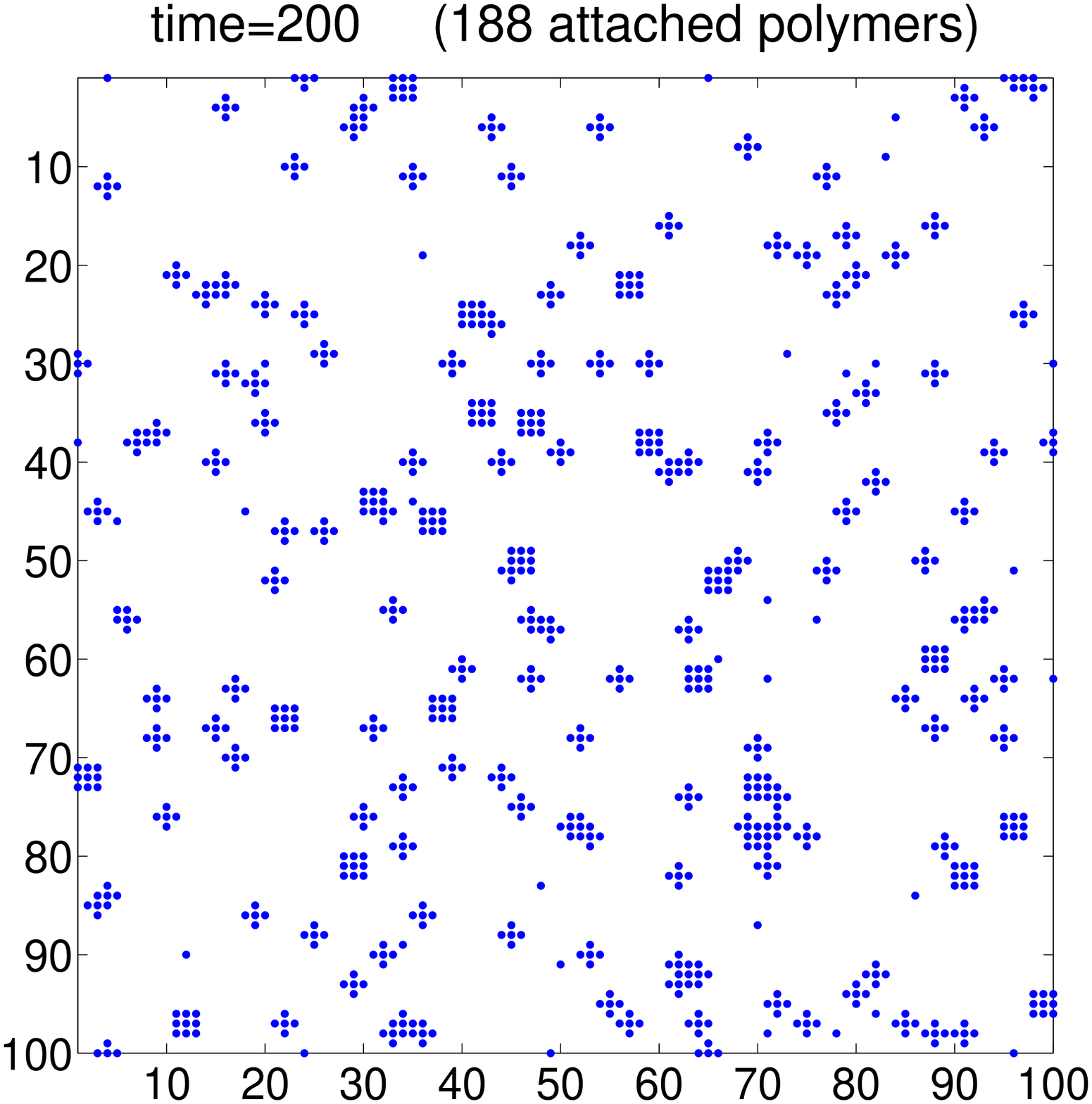,height=2.6in}
\!\!\!\!\!\!\!\!\!\!\!\!\!\!\!\!\!\!\!\!
\psfig{file=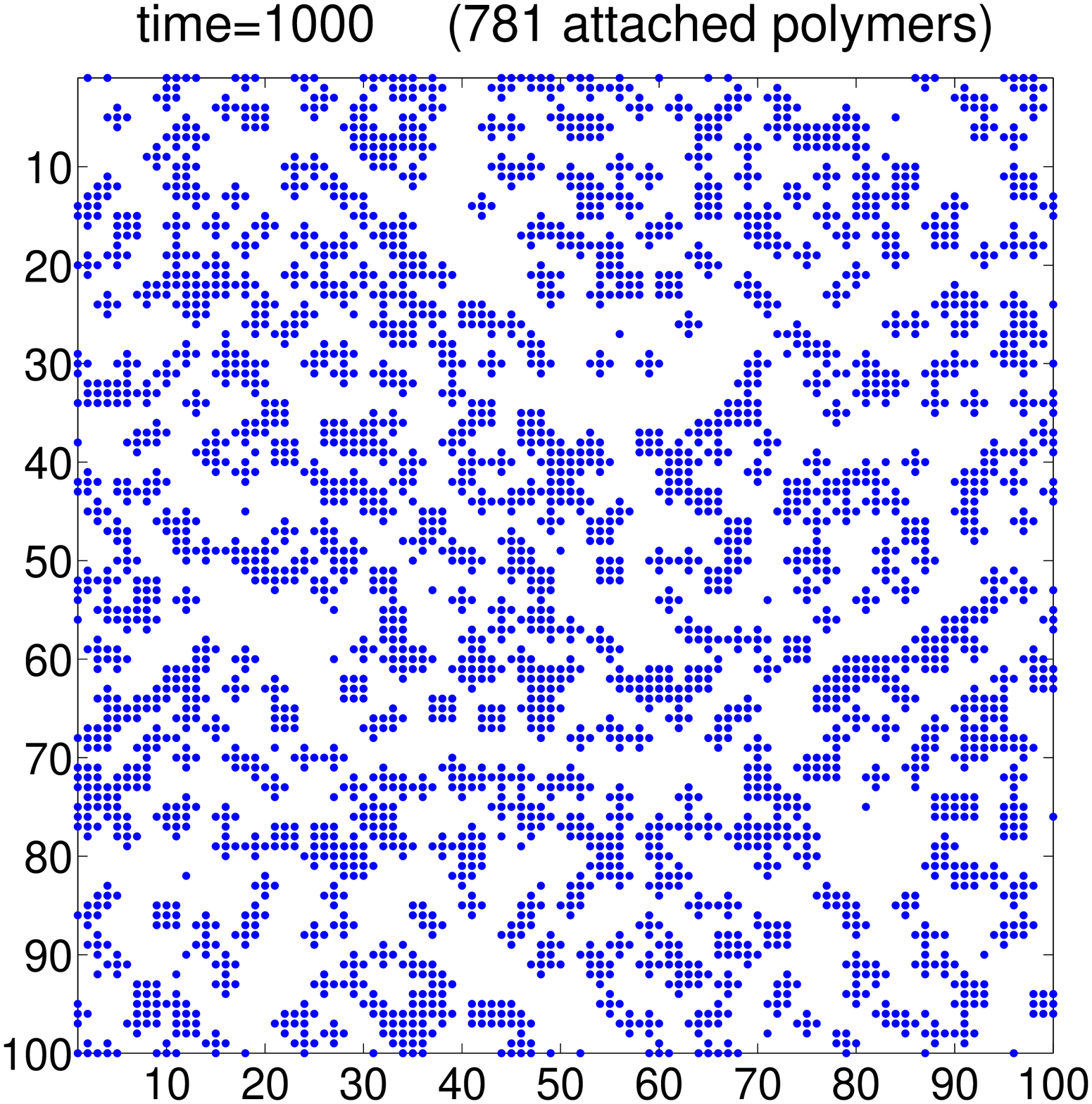,height=2.6in}
}
\centerline{
\psfig{file=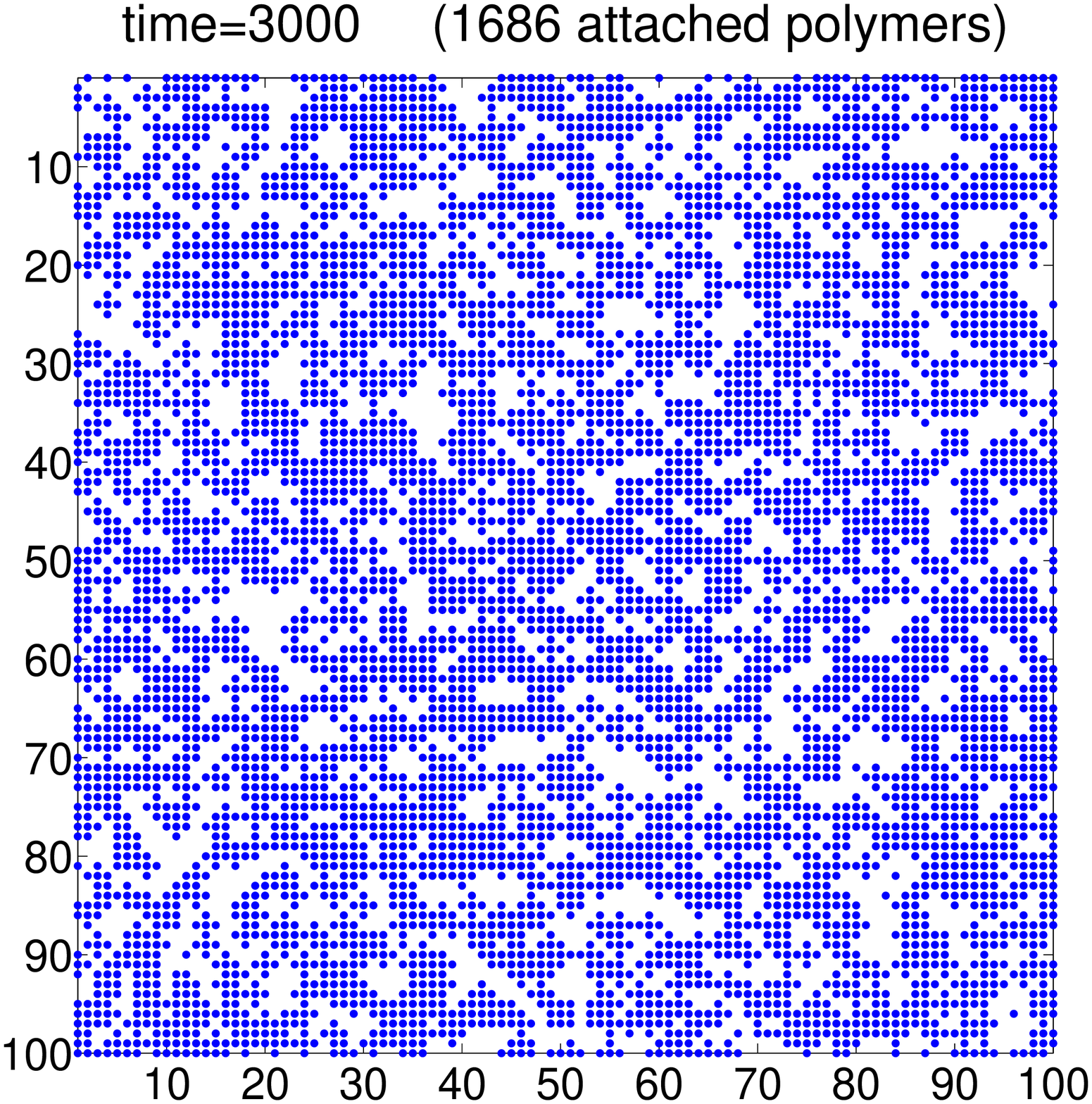,height=2.6in}
\!\!\!\!\!\!\!\!\!\!\!\!\!\!\!\!\!\!\!\!
\psfig{file=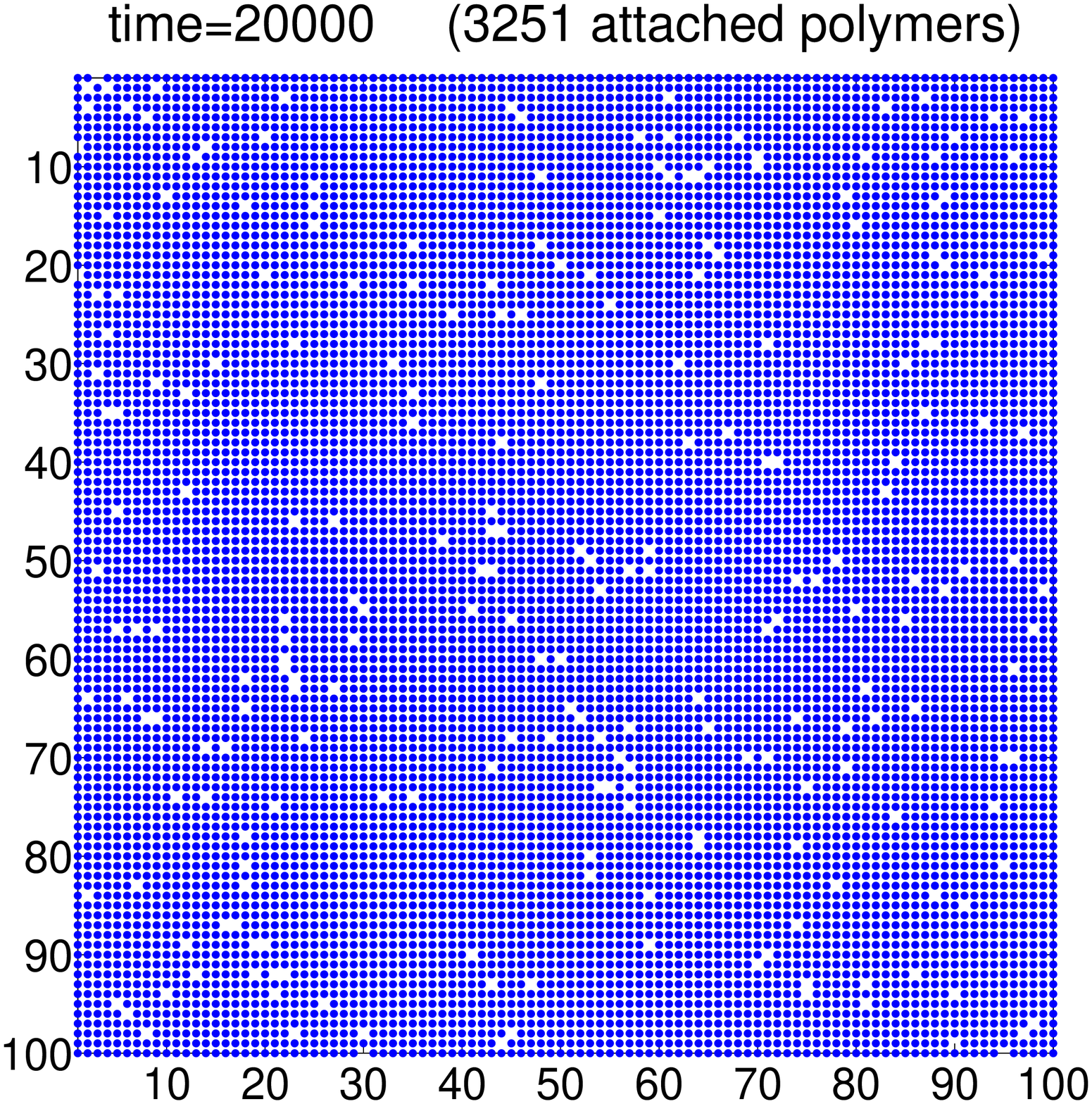,height=2.6in}
}
\vskip -5mm
\caption{{\it One realization of pRSA algorithm for $\alpha=0.8$ 
and $\beta=0.1$. The covered mesh points of the rectangular 
$100 \times 100$ mesh are shown at different times.
}}
\label{figal2surfplots0801}
\end{figure}
is shown in Figure \ref{figal2surfplots0801}. We start
with an empty rectangular $100 \times 100$ mesh,
i.e. $M=100.$ The mesh points covered by polymers
are plotted at different times. 

Let us note that pRSA algorithm requires that the position 
$(i,j)$ of the center of the adsorbed cross $\cros_{i,j}$ (resp. square 
$\squ_{i,j}$) is vacant. On the other hand, the ``tails" of crosses/squares 
can overlap. Here, the center of the cross (resp. square) describes the 
reactive group which is covalently bound to the surface. The remaining four 
(resp. eight) points of the cross (resp. square) describe the polymer tails 
which sterically shield the neighbourhood of the adsorbed polymer. In our 
algorithm, binding of a larger polymer prevents  binding (of the center) 
of another 
polymer in the neighbourhood of the center of the polymer already 
adsorbed. On the other hand, the ``wiggling  tails" of polymers can overlap. 

As in \cite{Erban:2006:DPI} there are two important quantities of interest: 
the number of covered mesh points $A(t)$ and the number $N(t)$ of polymers 
 which are attached to the surface 
at time $t$. To understand the behaviour of $A(t)$ and $N(t)$,
we introduce in Section \ref{secoperformalism} an operator formalism 
which makes it possible to derive a series expansion for $N(t)$. 
We also derive series for $A(t)$ and for numbers
of point-polymers, cross-polymers and square-polymers adsorbed
on the surface at time $t$. The operator formalism transforms
the random sequential adsorption process into a set of combinatorial
problems on the lattice. In some special cases, one can further simplify 
the resulting lattice combinatorial problems; we consider these
special cases in Section \ref{secspecialcases}. The general problem is studied
in Section \ref{secgeneralcase}. To illustrate the precision
of the derived formulas, we also provide a comparison of the results
obtained by series expansion with those obtained by direct stochastic 
simulation, of particular interest is the time evolution of $A(t)$ and $N(t)$ 
and the dependence of the final adsorbed polymer layer on the
parameters $\alpha$ and $\beta$. We conclude with a discussion 
in Section \ref{secdiscussion}.

\section{Operator formalism}

\label{secoperformalism}

Let us denote by $N(t)$ (resp. $N_p(t)$, $N_c(t)$ and $N_s(t)$)
the number of polymers (resp. point-polymers, cross-polymers 
and square-polymers) which are adsorbed on the surface
at time $t$. Then we have
\begin{equation}
N_p(t) = (1 - \alpha - \beta) N(t),
\qquad
N_c(t) = \alpha N(t),
\qquad
N_s(t) = \beta N(t).
\label{obvrelations}
\end{equation}
Let $A(t)$ (resp. $F(t)$) be the number of covered
(resp. vacant) mesh points at time $t$. 
Since $dN/dt = F/M^2$ and $A = M^2 - F$, we have 
\begin{equation}
A(t) = M^2 \left(1 - \frac{\dd N}{\dt}(t) \right).
\label{obvrelations2}
\end{equation}
Let us define
\begin{equation}
N^\infty = \lim_{t \to \infty} N(t),
\quad
N_p^\infty = \lim_{t \to \infty} N_p(t),
\quad
N_c^\infty = \lim_{t \to \infty} N_c(t),
\quad
N_s^\infty = \lim_{t \to \infty} N_s(t).
\label{limitdef}
\end{equation}
Then (\ref{obvrelations}) implies
\begin{equation}
N_p^\infty = (1 - \alpha - \beta) N^\infty,
\qquad
N_c^\infty = \alpha N^\infty,
\qquad
N_s^\infty = \beta N^\infty.
\label{obvrelations3}
\end{equation}
Hence, the saturating values $N_p^\infty,$ $N_c^\infty$ and $N_s^\infty$
can be computed directly from $N^\infty$. Similarly, the time evolution
of $A(t)$, $N_p(t)$, $N_c(t)$ and $N_s(t)$ can be obtained from $N(t)$
by (\ref{obvrelations}) -- (\ref{obvrelations2}). In this section,
we develop an operator formalism framework to obtain the time evolution 
of $N(t)$ and the limit $N^\infty$. Once we get $N(t)$ and $N^\infty$, 
the rest of quantities of interest can be expressed by 
(\ref{obvrelations}), (\ref{obvrelations2}) and (\ref{obvrelations3})
and their dependence on the model parameters $\alpha$ and $\beta$
can be also studied.

In \cite{Dickman:1991:RSA,Fan:1991:ACR}, an operator formalism was developed
for studying the square lattice with nearest-neighbour exclusion.
The results can be directly used to find an approximation
of $N^\infty$ for $\alpha=1$ and $\beta=0$. If $\alpha=1$, then it is 
sufficient to keep track of the centers of cross-polymers. Each center
of a cross-polymer excludes putting another center of a
cross-polymer in the nearest neighbourhood of it. Hence, one
can reformulate pRSA algorithm for $\alpha=1$ in terms of 
adsorption of points which excludes the nearest neighbourhood
of them. Similarly, one can reformulate the pRSA algorithm as
adsorption of points which excludes the nearest and the next
nearest neighbourhood of them for $\alpha=0$ and $\beta = 1$.
However, if  $[\alpha,\beta] \not \in \{[0,1],[1,0], [0,0] \}$, 
then we have a mixture of polymers of different sizes in the
solution and the approach of \cite{Dickman:1991:RSA,Fan:1991:ACR} 
cannot be directly used. In this section we present a generalization
of the operator formalism for the case of arbitrary $\alpha$
and $\beta$.

We consider an $M \times M$ lattice (with periodic boundary
conditions) to which polymers can adsorb. 
For each lattice point $(i,j)$, we consider the state function
$\nu_{i,j} \in \{0, 1, 2, 3\}.$ Here, $\nu_{i,j}=0$ means that
lattice point $(i,j)$ is vacant or occupied by the ``wiggling tail" of
a cross-polymer/square-polymer (i.e. $\nu_{i,j}=0$ means that
lattice point $(i,j)$ is free of centers of polymers/attached 
reactive groups),
$\nu_{i,j}=1$ means that lattice point $(i,j)$ is occupied by the
point-polymer, $\nu_{i,j}=2$ means that the lattice point 
is occupied by the center of the cross-polymer and
$\nu_{i,j}=3$ means that the lattice point is occupied by the center 
of the square-polymer. Denoting
$$
|0\rangle = [1,0,0,0]^T\!\!,
\qquad
|1\rangle = [0,1,0,0]^T\!\!,
\qquad
|2\rangle = [0,0,1,0]^T\!\!,
\qquad
|3\rangle = [0,0,0,1]^T\!\!,
$$
we identify every lattice point with the four-dimensional vector space 
$\er^4$. Namely, the configuration of the $M \times M$ lattice will 
be expressed as
$$ 
| \{ \nu_{i,j} \} \rangle
\in
\mathop{\bigoplus}_{i,j=1}^{M} 
 \Big\{ |0\rangle,|1\rangle,|2\rangle,|3\rangle \Big\}.
$$
The system state is given by 
\begin{equation}
|\Psi(t)\rangle
=
\sum_{\{ \nu_{i,j} \}} P(\{ \nu_{i,j} \},t) | \{ \nu_{i,j} \} \rangle
\label{ensemsysstates}
\end{equation}
where the sum is taken over all possible configurations $\{ \nu_{i,j} \}$
of the lattice and $P(\{ \nu_{i,j} \},t)$ is the probability of each
configuration. It satisfies the normalization condition
\begin{equation}
\sum_{\{ \nu_{i,j} \}} P(\{ \nu_{i,j} \},t) = 1.
\label{normprob}
\end{equation}
For each lattice point, we define cross, square and point annihilation 
operators
\begin{equation}
\boldA 
=
\left(
\begin{matrix}
0 & 0 & 1 & 0 \\
0 & 0 & 0 & 0 \\
0 & 0 & 0 & 0 \\
0 & 0 & 0 & 0
\end{matrix}
\right),
\qquad
\boldB 
=
\left(
\begin{matrix}
0 & 0 & 0 & 1 \\
0 & 0 & 0 & 0 \\
0 & 0 & 0 & 0 \\
0 & 0 & 0 & 0
\end{matrix}
\right),
\qquad
\boldC 
=
\left(
\begin{matrix}
0 & 1 & 0 & 0 \\
0 & 0 & 0 & 0 \\
0 & 0 & 0 & 0 \\
0 & 0 & 0 & 0
\end{matrix}
\right).
\label{anihcreat}
\end{equation}
More precisely, operator $\boldA_{i,j}$ (resp. $\boldB_{i,j}$, 
and $\boldC_{i,j}$) acts as $\boldA$ (resp. $\boldB$, and $\boldC$) 
on the lattice point $(i,j)$ and as the identity on all other lattice points.
We also define creation operator $\boldA_{i,j}^\dagger$ 
(resp. $\boldB_{i,j}^\dagger$, and $\boldC_{i,j}^\dagger$) 
as the transpose of operator $\boldA_{i,j}$ 
(resp. $\boldB_{i,j}$, and $\boldC_{i,j}$). 
The cross-polymer number operator (resp. square-polymer number operator, 
and point-polymer number operator)
is defined as $\boldN^c_{i,j} = \boldA_{i,j}^\dagger \boldA_{i,j}$
(resp. $\boldN^s_{i,j} = \boldB_{i,j}^\dagger \boldB_{i,j}$,
and $\boldN^p_{i,j} = \boldC_{i,j}^\dagger \boldC_{i,j}$)
which is, at the lattice point $(i,j)$, a projection onto the
one-dimensional subspace spanned by vector $|2\rangle$ which
corresponds to a cross. 
The ``vacancy" number operator can be expressed as
$\boldN^v_{i,j} = \boldA_{i,j} \boldA_{i,j}^\dagger
= \boldB_{i,j} \boldB_{i,j}^\dagger = \boldC_{i,j} \boldC_{i,j}^\dagger$. 
Here, ``vacancy" means that the lattice point is either free or
covered by the tail of the cross-polymer/square-polymer, i.e.
it is free of the attached reactive groups. We have
$\boldN^v_{i,j} \, | \Psi \rangle = 
(1 - \nu_{i,j})(2 - \nu_{i,j})(3 - \nu_{i,j})/6 \, | \Psi \rangle.$
Let $\boldR_{i,j}$ be an operator which is equal to the identity $Id$ 
(resp. $0$) operating on configurations in which lattice point
$(i,j)$ lies outside (within) the set of lattice points
covered by tails of cross-polymers or square-polymers, i.e.
\begin{eqnarray}
\boldR_{i,j}
\! \! \! &
\equiv
& \! \! \! 
(\boldN^v_{i+1,j} + \boldN^p_{i+1,j})
(\boldN^v_{i-1,j} + \boldN^p_{i-1,j})
(\boldN^v_{i,j+1} + \boldN^p_{i,j+1})
(\boldN^v_{i,j-1} + \boldN^p_{i,j-1})
\nonumber
\\
& &
\! \! \! \circ
(\boldN^v_{i+1,j+1} + \boldN^p_{i+1,j+1} + \boldN^c_{i+1,j+1})
(\boldN^v_{i-1,j+1} + \boldN^p_{i-1,j+1} + \boldN^c_{i-1,j+1})
\qquad
\label{operR}
\\
& &
\! \! \! \circ
(\boldN^v_{i+1,j-1} + \boldN^p_{i+1,j-1} + \boldN^c_{i+1,j-1})
(\boldN^v_{i-1,j-1} + \boldN^p_{i-1,j-1} + \boldN^c_{i-1,j-1}),
\nonumber
\end{eqnarray}
where symbol $\circ$ is used to emphasize that $\boldR_{i,j}$ is the 
composition of operators which are typed on several lines
(to simplify the resulting formulas, we skip the composition symbol 
$\circ$ if the composed operators are typed on the same line). 
In the pRSA algorithm, we add the center of a cross-polymer 
(resp. square-polymer, and point-poly\-mer) at the lattice 
site $(i,j)$ at a rate $\alpha$ (resp. $\beta$, 
and $1-\alpha - \beta$) if $(i,j)$ is vacant and not covered by the tail 
of a cross-polymer or square-polymer.
This means that the state $|\Psi(t) \rangle$ satisfies the master
equation
\begin{equation}
\frac{\partial}{\partial t}
| \Psi(t) \rangle
=
\frac{1}{M^2}
\sum_{i,j=1}^M
\Big(
\alpha
\boldA^\dagger_{i,j}
\boldR_{i,j}
+
\beta
\boldB^\dagger_{i,j}
\boldR_{i,j}
+
(1 - \alpha - \beta)
\boldC^\dagger_{i,j}
\boldR_{i,j}
-
\boldN^v_{i,j} 
\boldR_{i,j}
\Big)
| \Psi(t) \rangle.
\label{mastereq}
\end{equation}
Solving (\ref{mastereq}) with the initial condition
$|\{0\}\rangle = |0\rangle \oplus |0\rangle \oplus \dots \oplus |0\rangle$, 
we obtain
\begin{equation}
|\Psi(t) \rangle
=
\exp
\Bigg[
\frac{t}{M^2}
\sum_{i,j=1}^M
\Big(
\alpha
\boldA^\dagger_{i,j}
+
\beta
\boldB^\dagger_{i,j}
+
(1 - \alpha - \beta)
\boldC^\dagger_{i,j}
-
\boldN^v_{i,j} 
\Big)
\boldR_{i,j}
\Bigg]
 |\{0\}\rangle.
\label{solmastereq}
\end{equation}
Denoting 
$|\{u\}\rangle = |u\rangle \oplus |u\rangle \oplus \dots \oplus |u\rangle$
where $|u\rangle = [1,1,1,1]^T$,
we can compute the number of polymers at time $t$ by
\begin{equation}
N(t) 
=
M^2 \langle \{u\} | 
(\boldN^p_{1,1} + \boldN^c_{1,1} + \boldN^s_{1,1}) | \Psi(t) \rangle.
\label{formNtgen}
\end{equation}
Using (\ref{solmastereq}), we get
$$
N(t) 
=
M^2 \Bigg\langle \{u\} \Bigg| 
(\boldN^p_{1,1} + \boldN^c_{1,1} + \boldN^s_{1,1})
\qquad \qquad  \qquad \qquad \qquad \qquad \qquad \qquad \qquad \qquad
$$
$$
\qquad \qquad \circ
\exp
\Bigg[
\frac{t}{M^2}
\sum_{i,j=1}^M
\Big(
\alpha
\boldA^\dagger_{i,j}
+
\beta
\boldB^\dagger_{i,j}
+
(1 - \alpha - \beta)
\boldC^\dagger_{i,j}
-
\boldN^v_{i,j} 
\Big)
\boldR_{i,j}
\Bigg]
\Bigg| \{0\} 
\Bigg\rangle
=
$$
$$
=
M^2 \sum_{k=0}^\infty
\frac{1}{k!} \left( \frac{t}{M^2} \right)^k 
\Bigg\langle \{u\} \Bigg| 
(\boldN^p_{1,1} + \boldN^c_{1,1} + \boldN^s_{1,1})
\qquad \qquad \qquad \qquad \qquad \qquad \qquad \; \;
$$
$$
\qquad \qquad \circ
\Bigg[
\sum_{i,j=1}^M
\Big(
\alpha
\boldA^\dagger_{i,j}
+
\beta
\boldB^\dagger_{i,j}
+
(1 - \alpha - \beta)
\boldC^\dagger_{i,j}
-
\boldN^v_{i,j} 
\Big)
\boldR_{i,j}
\Bigg]^k
\Bigg| \{0\} 
\Bigg\rangle
=
$$
\begin{equation}
=
M^2 
\sum_{k=0}^\infty
\frac{1}{k!} \left( \frac{t}{M^2} \right)^k 
\Bigg\langle \! \{u\} \Bigg| 
(\boldN^p_{1,1} + \boldN^c_{1,1} + \boldN^s_{1,1})
\qquad \qquad  \qquad \qquad \qquad \qquad
\label{formulaNt}
\end{equation}
$$
\qquad \qquad \circ \Bigg[
\sum_{\{(x_j,y_j)\}_{j=1}^k}
\prod_{j=1}^k
\Bigg[
\Big(
\alpha
\boldA^\dagger_{x_j,y_j}
+
\beta
\boldB^\dagger_{x_j,y_j}
+
(1 - \alpha - \beta)
\boldC^\dagger_{x_j,y_j}
-
\boldN^v_{x_j,y_j} 
\Big)
\boldR_{x_j,y_j}
\Bigg]
\Bigg| \{0\} 
\! \Bigg\rangle.
$$
Here, the last sum is done over all $k$-tuples
$\{(x_j,y_j)\}_{j=1}^k$ in the mesh. To evaluate
this formula, let us note that we can consider
the contributions of each mesh point separately. 
If $(x_i,y_i) \ne (1,1)$, then an operator of
the following type acts on the mesh point 
$(x_i,y_i)$:
$$
[
\boldN^v_{x_i,y_i}\! + \boldN^p_{x_i,y_i}
]^{\gamma_1}
[
\boldN^v_{x_i,y_i}\! + \boldN^p_{x_i,y_i}\! + \boldN^c_{x_i,y_i}
]^{\gamma_2}
[ 
\alpha \boldA^\dagger_{x_i,y_i}\!
+
\beta
\boldB^\dagger_{x_i,y_i}\!
+
(1 - \alpha - \beta)
\boldC^\dagger_{x_i,y_i}\!
-
\boldN^v_{x_i,y_i} 
]^{\gamma_3}
$$
\begin{equation}
\circ
[
\boldN^v_{x_i,y_i} + \boldN^p_{x_i,y_i}
]^{\gamma_4}
[
\boldN^v_{x_i,y_i} + \boldN^p_{x_i,y_i} + \boldN^c_{x_i,y_i}
]^{\gamma_5}
\dots
\equiv
\boldW_{x_i,y_i},
\label{buildblock}
\end{equation}
where $\gamma_1,$ $\gamma_2$, $\gamma_3$, \dots, are nonnegative integers.
Without loss of generality, we can assume $\gamma_3 > 0$ in what follows.
The ``building blocks" $\boldW_{x_i,y_i}$ can be reasonably 
simplified if we take into account the following formulas:
\begin{eqnarray} 
[ \boldN^v_{x_i,y_i} + \boldN^p_{x_i,y_i} ]^2
& = &
\boldN^v_{x_i,y_i} + \boldN^p_{x_i,y_i},
\nonumber
\\ \relax 
[\boldN^v_{x_i,y_i}+\boldN^p_{x_i,y_i}+\boldN^c_{x_i,y_i}]^2
& = &
\boldN^v_{x_i,y_i} + \boldN^p_{x_i,y_i} + \boldN^c_{x_i,y_i},
\nonumber
\\ \relax 
[ \boldN^v_{x_i,y_i} + \boldN^p_{x_i,y_i} ]
[ \boldN^v_{x_i,y_i} + \boldN^p_{x_i,y_i} + \boldN^c_{x_i,y_i} ]
& = &
\boldN^v_{x_i,y_i} + \boldN^p_{x_i,y_i},
\nonumber
\end{eqnarray}
\begin{eqnarray} 
[ 
\alpha \boldA^\dagger_{x_i,y_i}\!
+
\beta
\boldB^\dagger_{x_i,y_i}\!
+
(1 - \alpha - \beta)
\boldC^\dagger_{x_i,y_i}\!
- \!\!\!\!\!&&\!\!\!\!\!\!
\boldN^v_{x_i,y_i} 
]^{2}
=
\nonumber
\\ \relax 
=
-
[ 
\alpha \boldA^\dagger_{x_i,y_i}\!
+
\beta
\boldB^\dagger_{x_i,y_i}\!
+ \!\!\!\!\!&&\!\!\!\!\!\!
(1 - \alpha - \beta)
\boldC^\dagger_{x_i,y_i}\!
-
\boldN^v_{x_i,y_i} 
]
\boldN^v_{x_i,y_i},
\nonumber
\end{eqnarray}
\begin{eqnarray} 
[
\boldN^v_{x_i,y_i} + \boldN^p_{x_i,y_i}
]
[ 
\alpha \boldA^\dagger_{x_i,y_i}\!
+
\beta
\boldB^\dagger_{x_i,y_i}\!
+
(1 - \alpha - \beta)
\boldC^\dagger_{x_i,y_i}\!
- \!\!\!\!\!&&\!\!\!\!\!\!
\boldN^v_{x_i,y_i} 
]
=
\nonumber
\\ \relax 
=
[ 
(1 - \alpha - \beta) \!\!\!\!\!&&\!\!\!\!\!\! \boldC^\dagger_{x_i,y_i}
-
\boldN^v_{x_i,y_i} 
],
\nonumber
\end{eqnarray}
\begin{eqnarray} 
[
\boldN^v_{x_i,y_i} + \boldN^p_{x_i,y_i} + \boldN^c_{x_i,y_i}
]
[ 
\alpha \boldA^\dagger_{x_i,y_i}\!
+
\beta
\boldB^\dagger_{x_i,y_i}\!
+
(1 - \alpha - \beta)
\boldC^\dagger_{x_i,y_i}\!
- \!\!\!\!\!&&\!\!\!\!\!\!
\boldN^v_{x_i,y_i} 
]
=
\nonumber
\\ \relax 
=
[ 
\alpha \boldA^\dagger_{x_i,y_i}\!
+
(1 - \alpha - \beta) \!\!\!\!\!&&\!\!\!\!\!\! \boldC^\dagger_{x_i,y_i}
-
\boldN^v_{x_i,y_i} 
].
\nonumber
\end{eqnarray}
If $\gamma_1 = \gamma_2 =0 $, then the building block (\ref{buildblock}) can 
be rewritten in the form $\pm [\alpha \boldA^\dagger_{x_i,y_i} 
+ \beta \boldB^\dagger_{x_i,y_i}\! + (1 - \alpha - \beta)
\boldC^\dagger_{x_i,y_i}\! - \boldN^v_{x_i,y_i} ]
[\boldN^v_{x_i,x_i}]^\gamma$. We can easily observe that 
$$
\left\langle \{u\} \Bigg|
\pm [ 
\alpha \boldA^\dagger_{x_i,y_i}
+ 
\beta \boldB^\dagger_{x_i,y_i}\! 
+ 
(1 - \alpha - \beta) \boldC^\dagger_{x_i,y_i}\!
-
\boldN^v_{x_i,y_i} 
]
[
\boldN^v_{x_i,y_i}
]^\gamma
\Bigg| \{0\} 
\right\rangle
= 0.
$$
Consequently, the first necessary condition for $k$-tuple
$\{(x_j,y_j)\}_{j=1}^k$ to have nonzero contribution
to the formula (\ref{formulaNt}) is that for every
$(x_i,y_i) \ne (1,1)$ in the $k$-tuple, there must
be $j<i$ such that $(x_i,y_i)$ is equal to
$(x_j,y_j)$ or one of its nearest or next nearest neighbours.
In particular, we see that $(x_1,y_1) = (1,1)$
in order to have nonzero contribution of the $k$-tuple
$\{(x_j,y_j)\}_{j=1}^k$. If $\gamma_1 > 0$, then
(\ref{buildblock}) satisfies 
\begin{equation}
\left\langle \{u\} 
\Bigg|
\boldW_{x_i,y_i} 
\Bigg| \{0\} 
\right\rangle
= (-1)^{\overline{\gamma_i}} (\alpha+\beta)
\label{buildblocksum}
\end{equation}
where we have denoted by ${\overline{\gamma_i}}$ the number of times
that the mesh point $(x_i,y_i)$ appears in the $k$-tuple
$\{(x_j,y_j)\}_{j=1}^k$. Similarly, if $\gamma_1 = 0$
and $\gamma_2 > 0$, then
(\ref{buildblock}) satisfies 
\begin{equation}
\left\langle \{u\} 
\Bigg|
\boldW_{x_i,y_i} 
\Bigg| \{0\} 
\right\rangle
= (-1)^{\overline{\gamma_i}} \beta.
\label{buildblocksumb}
\end{equation}
Finally, considering the contribution
of the first mesh point $(x_1,y_1) = (1,1)$, we get
\begin{equation}
\left\langle\! \{u\} \Bigg| 
(\boldN^p_{1,1}\! + \!\boldN^c_{1,1}\! + \!\boldN^s_{1,1})
[ 
\alpha \boldA^\dagger_{1,1}\!
+
\!\beta \boldB^\dagger_{1,1}\! 
+ 
\!(1 - \alpha - \beta) \boldC^\dagger_{1,1}\!
-
\!\boldN^v_{1,1} 
]
\dots
\Bigg| \{0\} 
\!\right\rangle
= (-1)^{\overline{\gamma_1}-1}.
\label{buildblocksum2}
\end{equation}

\smallskip

\noindent
Let us define $\setP_k$ as the set of all sequences 
$s \equiv \{(x_j,y_j)\}_{j=1}^k$, such that $(x_1,y_1) = (1,1)$ and for each 
$i \in \{2, \dots, k\}$ there exists $j < i$ such that 
$(x_i,y_i) \in \squ_{x_j,y_j}$, i.e. $(x_i,y_i)$ is equal to
$(x_j,y_j)$ or one of its nearest or next nearest neighbours.
Let us denote by $\omega(s)$ the number of distinct points in the 
sequence $s \in \setP_k.$ Let $\xi(s)$ be the number of distinct
points $(x_i,y_i) \in s,$ $(x_i,y_i) \ne (1,1)$,
satisfying that there exists $j < i$ such that 
$(x_i,y_i) \in \cros_{x_j,y_j}$, i.e. $(x_i,y_i)$ is equal to
$(x_j,y_j)$ or one of its nearest neighbours.
Then we can rewrite (\ref{formulaNt}) 
(using (\ref{buildblocksum}) -- (\ref{buildblocksum2})) as
\begin{equation}
N(t)
=
M^2 
\sum_{k=1}^\infty
\frac{1}{k!} \left( \frac{t}{M^2} \right)^k 
(-1)^{k-1}
\sum_{s \in \setP_k}
[\alpha + \beta]^{\xi(s)} 
\beta^{\omega(s) - \xi(s) - 1}.
\label{Ntseries}
\end{equation}
Formula (\ref{Ntseries}) is a starting point for the analysis of 
the pRSA algorithm. In order to evaluate coefficients of the series 
expansion (\ref{Ntseries}), we have to compute the quantities 
\begin{equation}
\sum_{s \in \setP_k}
[\alpha + \beta]^{\xi(s)} \beta^{\omega(s) - \xi(s) - 1},
\qquad
\mbox{for} \; k=1,2,3,\dots.
\label{combquan}
\end{equation}
Thus we have transformed
the problem of the original pRSA algorithm to a combinatorial
problem on the two-dimensional lattice. The problem can be 
further simplified if $\alpha = 0$ or $\beta = 0$ as we will show
in the following section. The general analysis of 
(\ref{Ntseries}) for any $\alpha$ and $\beta$ is given in 
Section \ref{secgeneralcase}.

\section{Analysis of pRSA algorithm in some special cases}

\label{secspecialcases}

First, let us note that formula (\ref{Ntseries}) is
consistent with the trivial case $[\alpha, \beta] = [0,0]$.
We have
$$
N(t)  
=
M^2 
\sum_{k=1}^\infty
\frac{1}{k!} \left( \frac{t}{M^2} \right)^k 
(-1)^{k-1}
=
M^2
\left(
1 - 
\exp \left[ - \frac{t}{M^2} \right]
\right)
$$
which is the exact formula for $\alpha=\beta =0$. This can be seen
easily, since $N(t) \equiv A(t) = M^2 - F(t)$ where $F(t)$ solves
$dF/dt = - F/M^2$. Formula (\ref{Ntseries}) is
also consistent with the cases
$[\alpha, \beta]= [1,0]$ and $[\alpha, \beta]= [0,1]$
which were studied in \cite{Dickman:1991:RSA,Fan:1991:ACR}.
Choosing $[\alpha, \beta]= [0,1]$, we get 
\begin{equation}
N(t)  
=
M^2 
\sum_{k=1}^\infty
\frac{1}{k!} \left( \frac{t}{M^2} \right)^k 
(-1)^{k-1} \Big| \setP_k \Big|
\label{Ntcase01}
\end{equation}
which is the formula derived in \cite{Dickman:1991:RSA,Fan:1991:ACR}.
Here, $| \setP_k |$ is the number of sequences in $\setP_k$.
Similarly, if $[\alpha, \beta]= [1,0]$, we get 
\begin{equation}
N(t)  
=
M^2 
\sum_{k=1}^\infty
\frac{1}{k!} \left( \frac{t}{M^2} \right)^k 
(-1)^{k-1} \Big| \setQ_k \Big|
\label{Ntcase10}
\end{equation}
where $\setQ_k$ is the set of all sequences 
$s \equiv \{(x_j,y_j)\}_{j=1}^k$, such that $(x_1,y_1) = (1,1)$ 
and for each  $i \in \{2, \dots, k\}$ there exists $j < i$ such that 
$(x_i,y_i) \in \cros_{x_j,y_j}$, i.e. $(x_i,y_i)$ is equal to
$(x_j,y_j)$ or one of its nearest neighbours.

The situation is more complicated if 
$[\alpha, \beta] \not \in \{[0,0],[1,0],[0,1]\}.$
To evaluate coefficients of the series (\ref{Ntseries}),
we have to compute the quantities (\ref{combquan}). If we use 
directly formula (\ref{combquan}),
we would have to evaluate a different computationally intensive 
combinatorial problem for each $\alpha$ and $\beta$. Here we show
that we can transform formula
(\ref{Ntseries}) to the problem where computationally intensive
part (involving $\setP_k$ or $\setQ_k$) is done independently 
of $\alpha$ or $\beta$. We start with the analysis of the pRSA
algorithm in the special case $\alpha=0$.

\subsection{Special case $\alpha=0$}

\label{secalpha0}

\begin{figure}
\picturesAB{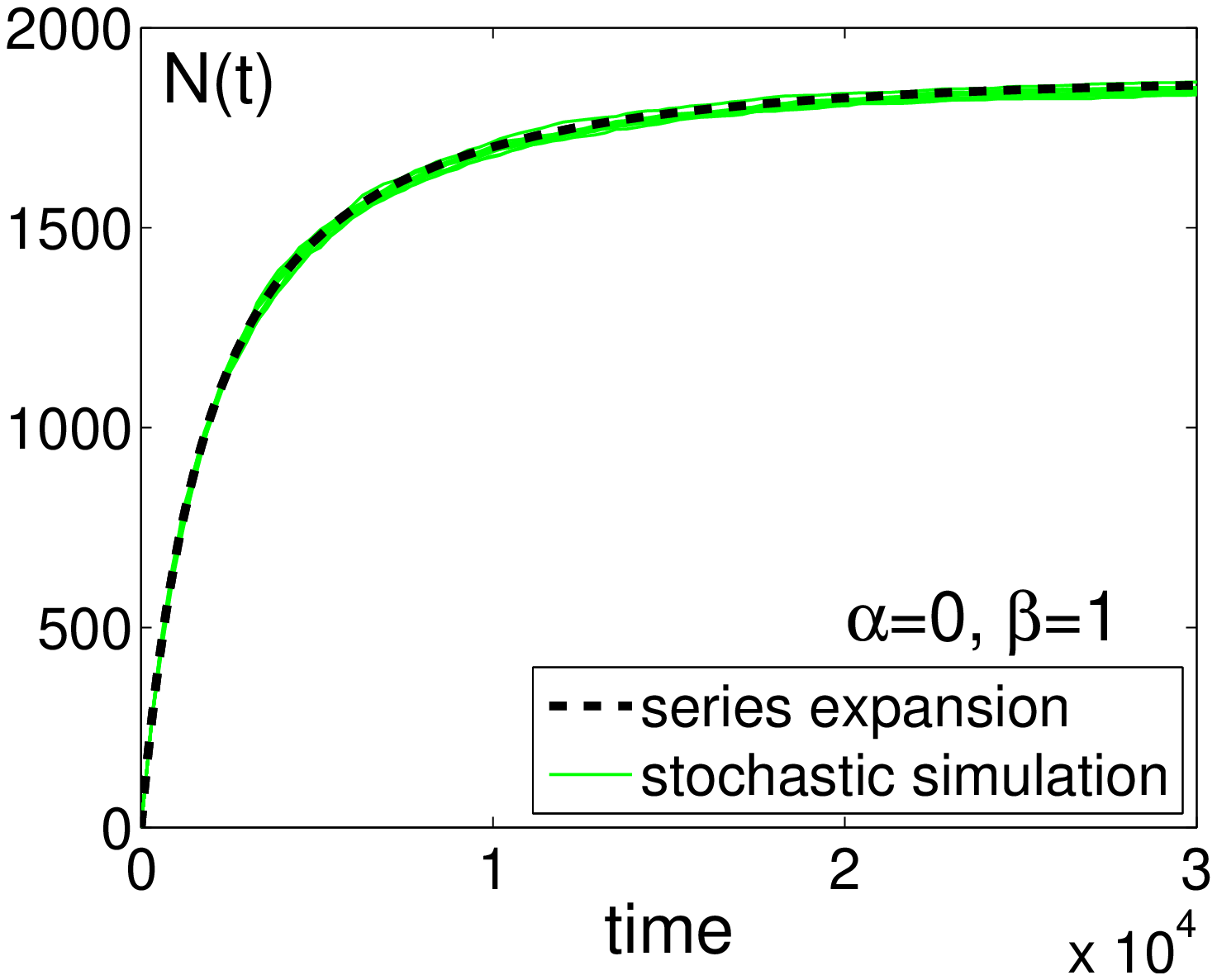}{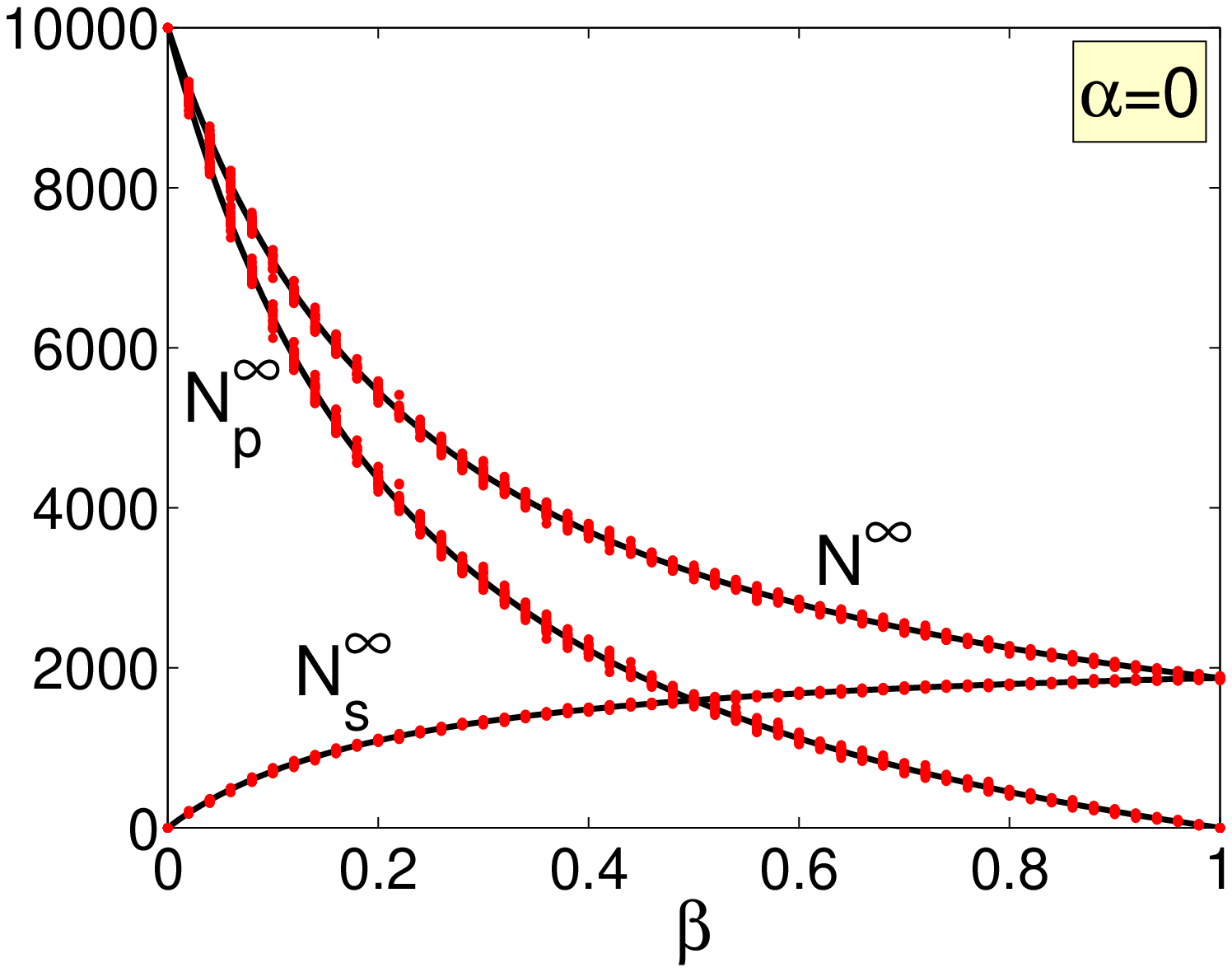}{2.07in}
\caption{{\it pRSA algorithm for $\alpha=0$.} 
(a) {\it Time evolution of $N(t)$ given by
$(\ref{approxNB})$ for $\alpha=0$, $\beta=1$ and $M=100$ (dashed line).
Ten realizations of stochastic simulation of pRSA algorithm are plotted
as thin solid lines.}
(b) {\it $N^\infty$, $N_p^\infty$ and $N_s^\infty$
as obtained by $(\ref{NNsNpinfty})$ (solid lines)
for $M=100$. We compare the approximate results
with stochastic simulation of pRSA algorithm (20 realizations,
each realization plotted as a dot).}}
\label{figseriesalpha0}
\end{figure}

If $\alpha=0$, then pRSA algorithm reduces to adsorption
of point-polymers and square-polymers, and (\ref{Ntseries}) reads 
as follows
\begin{equation}
N(t)
=
M^2 
\sum_{k=1}^\infty
\frac{1}{k!} \left( \frac{t}{M^2} \right)^k 
(-1)^{k-1}
\sum_{s \in \setP_k}
\beta^{\omega(s) - 1}.
\label{NtseriesB}
\end{equation}
If $\beta = 1$, then (\ref{NtseriesB}) implies (\ref{Ntcase01}). 
It was observed in \cite{Fan:1991:ACR} that the Laplace transform 
can be used to further simplify the formula (\ref{Ntcase01}). 
Here, we show that the Laplace transform can help us to analyse 
(\ref{NtseriesB}) for any $\beta$. Taking the Laplace transform of 
(\ref{NtseriesB}), term by term, we obtain 
\begin{equation}
\widehat{N}(u)
=
\int_0^\infty
N(t)
e^{-ut}
\dt
=
-
\frac{M^2}{u} 
\sum_{k=1}^\infty
 \left( - \frac{1}{u M^2} \right)^k 
\sum_{s \in \setP_k}
\beta^{\omega(s)-1}
\label{NtseriesLaplaceB}
\end{equation}
for sufficiently large $u$. Let us define $\setG_k$ as the set of all 
sequences of $k$ {\it distinct} points $\{(x_j,y_j)\}_{j=1}^k$, such that 
$(x_1,y_1) = (1,1)$, and for each $i \in \{2, \dots, k\}$ there exists 
$j < i$ such that $(x_i,y_i)$ is equal to one of the nearest or
the next nearest neighbours  of $(x_j,y_j)$, 
i.e. $(x_i,y_i) \in \squ_{x_j,y_j}$ and $(x_i,y_i) \ne (x_j,y_j)$. 
Then we have (using (\ref{NtseriesLaplaceB}))
$$
\widehat{N}(u)
=
-
\frac{M^2}{u} 
\sum_{k=1}^\infty
\left( - \frac{1}{u M^2} \right)^k 
\Big| \setG_k \Big| \beta^{k-1}
\prod_{j=1}^k
\left[
1 
+ 
\left( - \frac{j}{u M^2} \right)
+ 
\left( - \frac{j}{u M^2} \right)^2 
+ \dots 
\right]
=
$$
$$
=
\frac{M^2}{u} 
\sum_{k=1}^\infty
\! \left( - \beta \right)^{k-1} 
\Big| \setG_k \Big| 
\prod_{j=1}^k
(u M^2 + j)^{-1}
=
\frac{M^2}{u} 
\sum_{k=1}^\infty
\frac{\left( - \beta \right)^{k-1} \big|\setG_k\big|}{(k-1)!} 
\! \int_0^1 \! (1-x)^{u M^2} x^{k-1} \dx
=
$$
$$
=
\frac{M^2}{u} 
\int_0^1
(1-x)^{u M^2}
\sum_{k=1}^\infty
\frac{\left( - \beta x \right)^{k-1}}{(k-1)!} 
\Big| \setG_k \Big| 
\dx.
$$
Taking the inverse Laplace transform, we obtain
(for sufficiently large $r$)
\begin{equation}
N(t)
=
\frac{1}{2 \pi} \int_{-\infty}^{\infty}
\widehat{N}(r + i u) e^{(r + i u) t} \du
=
M^2
\int_0^{1-\exp[-t/M^2]} 
\sum_{k=1}^\infty
\frac{\left( - \beta x \right)^{k-1}}{(k-1)!} 
\Big| \setG_k \Big| 
\dx.
\label{pomNtB}
\end{equation}
Let us define function 
\begin{equation}
\Psi(x) = M^2
\sum_{k=1}^\infty
\frac{\left( - \beta \right)^{k-1} x^k}{k!} 
\Big| \setG_k \Big|. 
\label{funpsi}
\end{equation}
Then (\ref{pomNtB}) yields
\begin{equation}
N(t) = \Psi \left( 1 - \exp \left[ - \frac{t}{M^2} \right] \right).
\label{Npsi}
\end{equation}
In particular, the final coverage of the lattice can be computed as
\begin{equation}
N^\infty \equiv \lim_{t \to \infty} N(t) = \lim_{x \to 1} \Psi(x)
\label{asympcovB}
\end{equation}
and other quantities of interest can be obtained by
(\ref{obvrelations}), (\ref{obvrelations2}) and (\ref{obvrelations3}).
Thus, the adsorption algorithm has been reformulated to the problem
of finding the numbers of sequences in the sets $\setG_k$, $k=1,2, \dots.$
Once, we have the numbers $|\setG_k|$ we can write $\Psi(x)$ for
any $\beta$ and compute $N(t)$, $N_p(t)$, $N_s(t)$ and $A(t)$ 
by (\ref{Npsi}), (\ref{obvrelations}) and (\ref{obvrelations2}), 
provided that we can compute the sum of series (\ref{funpsi})
with reasonable precision. To do so, we set
\begin{equation}
g_k =  M^2 \, \frac{\big| \setG_k \big|}{k!}.
\label{defgk}
\end{equation}
The first eight values of $g_k$ can be computed relatively easily as
follows $g_1 = 10 000;$ $g_2 = 40 000;$ 
$g_3 \doteq 146 667;$ $g_4 \doteq 500 000;$ 
$g_5 \doteq 1 606 667;$ $g_6 \doteq 4 918 889;$ 
$g_7 \doteq 14 461 429$ and $g_8 \doteq 41 070 290$.
Our task is to estimate the sum of series (\ref{funpsi}) 
knowing only the first eight partial sums
\begin{equation}
s_n (x) = \sum_{k=1}^n \left( - \beta \right)^{k-1} g_k x^k,
\qquad
n = 1, 2, \dots, 8.
\label{partialsumssk}
\end{equation}
To do that, we use Shanks transformation 
\cite{Shanks:1955:NTD} computed by Wynn's algorithm
\cite{Wynn:1956:DCT,Weniger:1989:NST} in the following way
\begin{eqnarray}
\varepsilon_1^n(x) & = & s_n(x),
\qquad
\mbox{for} \; n = 1, 2, \dots, 8,  
\nonumber
\\
\varepsilon_2^{n-1}(x) & = & 1/(s_n(x) - s_{n-1}(x))
\qquad
\mbox{for} \; n = 2, 3, \dots, 8, 
\nonumber
\\
\varepsilon_{k}^{n-k+1}(x) & = & \varepsilon_{k-2}^{n-k+2}(x)
+ 
1/( \varepsilon_{k-1}^{n-k+2}(x) - \varepsilon_{k-1}^{n-k+1}(x) )
\label{epsalgorithm}
\\
&&
\mbox{for} \; k = 3, 4, 5, 6, 7, \; n = k, \dots, 8. 
\nonumber
\end{eqnarray}
To approximate the sum $\Psi(x)$, we use the term $\varepsilon_7^2(x)$
which is also the Pad\'e $[4,3]$-approximant since we use Shanks transformation 
for a power series \cite{Weniger:1989:NST}. Thus, we aproximate
number of attached polymers as
\begin{equation}
N(t) \approx \varepsilon_7^2 \Big( x(t) \Big),
\; \;\;
\mbox{where} \;\;\; 
x(t) 
=
1-\exp \left[- \frac{t}{M^2} \right].
\label{approxNB}
\end{equation}
The results obtained by (\ref{approxNB}) for $\beta=1$ 
and $M=100$ are given in Figure \ref{figseriesalpha0}(a).
To compute the time evolution of $N(t)$ we chose an equidistant
mesh for $x$ in the interval $(0,1)$ and evaluated 
$\varepsilon_7^2$ by (\ref{epsalgorithm}) at each 
mesh point. Then the corresponding time $t$ was computed by (\ref{approxNB}).
In Figure \ref{figseriesalpha0}(a), we compare results obtained
by approximation (\ref{approxNB}) and by stochastic
simulation of pRSA algorithm. We see that we get an excellent
agreement between the theoretically derived formula
and the simulation. The asymptotic coverage can be approximated
as 
\begin{equation}
N^\infty \equiv \lim_{t \to \infty} N(t)
\approx 
\varepsilon_7^2(1),
\quad
N^\infty_s \approx \beta \,  
\varepsilon_7^2(1),
\quad
\mbox{and}
\quad
N^\infty_p \approx (1 - \beta) \, 
\varepsilon_7^2(1).
\label{NNsNpinfty}
\end{equation}
In Figure \ref{figseriesalpha0}(b), we compare  approximations
of $N^\infty$, $N_p^\infty$ and $N_s^\infty$ computed by 
(\ref{NNsNpinfty}) with the results
of stochastic simulations. %
We see that approximations (\ref{NNsNpinfty}) 
provide very good results for any $\beta$.
We can estimate relative error between approximation
$N_{app}^\infty$ and exact value $N^\infty$ as 
$(N_{app}^\infty-N^\infty)/N^\infty$. Using (\ref{NNsNpinfty}), 
we obtain $N_{app}^\infty - N^\infty \doteq 2.7$
and $(N_{app}^\infty-N^\infty)/N^\infty \doteq 0.15 \%$ 
for $\beta = 1$. Here, exact value of $N^\infty$ was approximated 
by averaging over 100,000 realizations of the pRSA algorithm as 
$N^\infty \doteq 1869.8$ for $\beta=1.$

\subsection{Special case $\beta = 0$}

\label{secbeta0}

\begin{figure}
\picturesAB{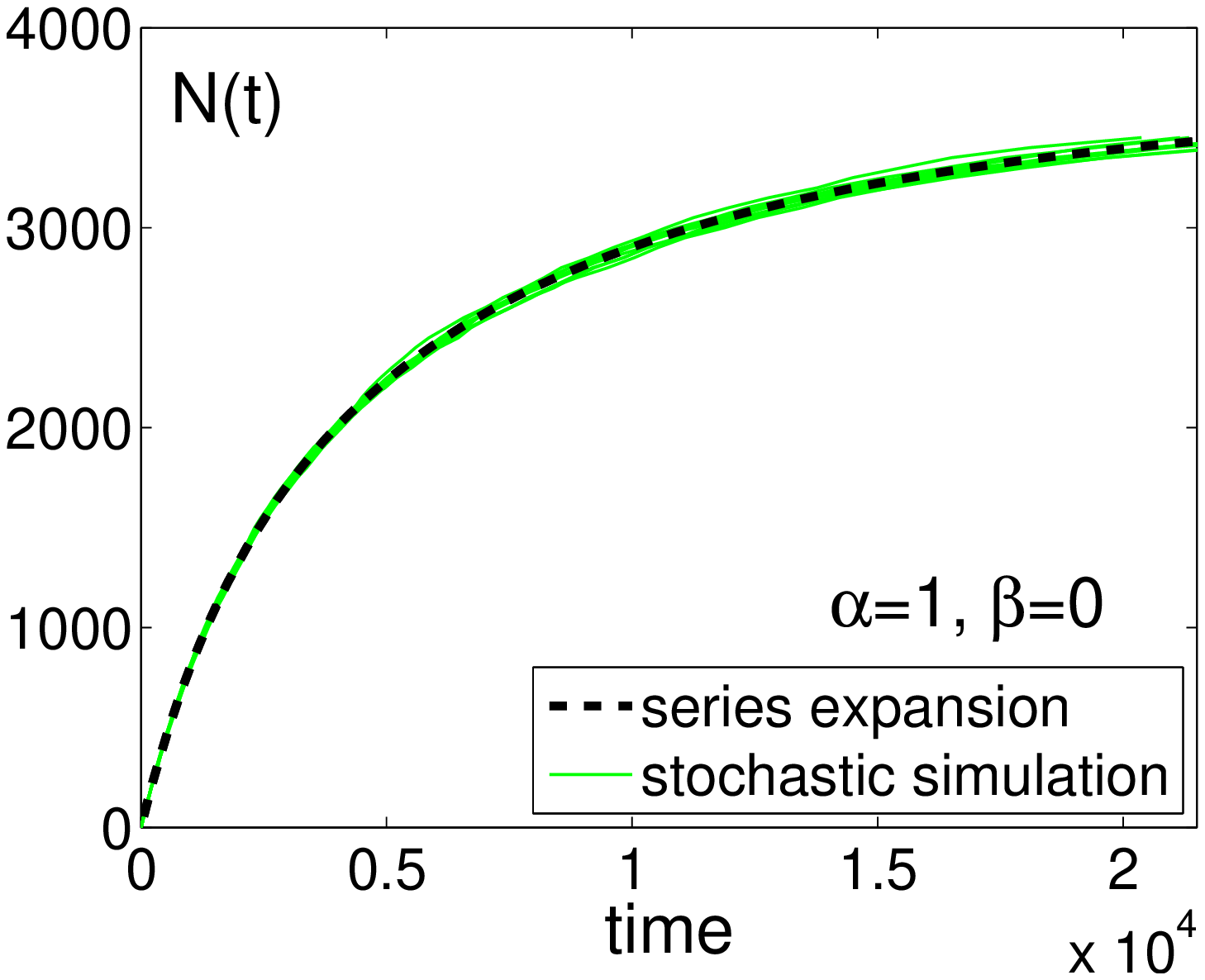}{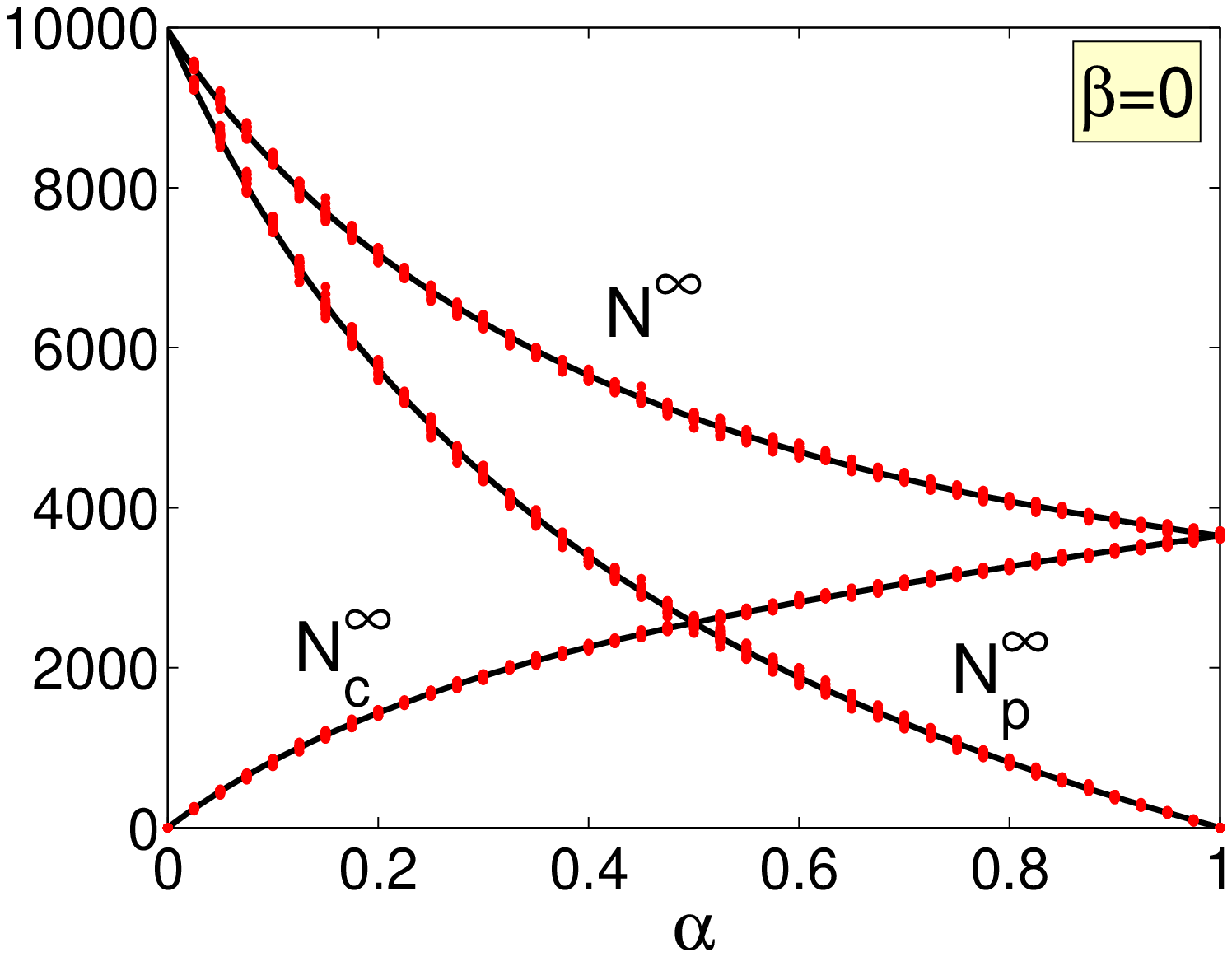}{2.07in}
\caption{{\it pRSA algorithm for $\beta=0$.} 
(a) {\it Time evolution of $N(t)$ given by
$(\ref{approxN})$ for $\alpha=1$, $\beta=0$ and $M=100$ (dashed line).
Ten realizations of stochastic simulation of pRSA algorithm are plotted
as thin solid lines.}
(b) {\it $N^\infty$, $N_p^\infty$ and $N_c^\infty$
as obtained by $(\ref{NNcNpinfty})$ (solid lines)
for $M=100$. We compare the approximate results
with stochastic simulation of pRSA algorithm (20 realizations,
each realization plotted as a dot).}}
\label{figseries}
\end{figure}

If $\beta=0$, then the pRSA algorithm reduces to the adsorption
of point-polymers and cross-polymers. The terms in the sum
(\ref{combquan}) are nonzero only if $\xi(s) = \omega(s) - 1$. 
Hence, (\ref{combquan}) can be rewritten as 
\begin{equation}
\sum_{s \in \setQ_k}
\alpha^{\omega(s) - 1},
\qquad
\mbox{for} \; k=1,2,3,\dots,
\label{combquan2}
\end{equation}
where $\setQ_k$ is the set of all sequences 
$s \equiv \{(x_j,y_j)\}_{j=1}^k$, such that $(x_1,y_1) = (1,1)$ 
and for each  $i \in \{2, \dots, k\}$ there exists $j < i$ such that 
$(x_i,y_i) \in \cros_{x_j,y_j}.$ As before,
$\omega(s)$ is the number of distinct points in the 
sequence $s \in \setQ_k.$ Using (\ref{combquan2}), 
we can rewrite (\ref{Ntseries}) as
\begin{equation}
N(t)
=
M^2 
\sum_{k=1}^\infty
\frac{1}{k!} \left( \frac{t}{M^2} \right)^k 
(-1)^{k-1}
\sum_{s \in \setQ_k}
\alpha^{\omega(s) - 1}.
\label{Ntseries2}
\end{equation}
Comparing formulas (\ref{NtseriesB}) and (\ref{Ntseries2}),
we find out only two differences: $\setP_k$ in (\ref{NtseriesB})
is replaced by $\setQ_k$ in (\ref{Ntseries2}) and
$\beta$ in (\ref{NtseriesB}) is replaced by $\alpha$
in (\ref{Ntseries2}). Consequently, taking the
Laplace transform of (\ref{Ntseries2}) and using the same method
as in Section \ref{secalpha0}, we find
(compare with (\ref{Npsi}))
\begin{equation}
N(t) = \Omega \left( 1 - \exp \left [- \frac{t}{M^2} \right] \right)
\quad
\mbox{where}
\quad
\Omega(x) = M^2
\sum_{k=1}^\infty
\frac{\left( - \alpha \right)^{k-1} x^k}{k!} 
\Big| \setH_k \Big|,
\label{Nomegafunomega}
\end{equation}
where $\setH_k$ is the set of all 
sequences of $k$ {\it distinct} points $\{(x_j,y_j)\}_{j=1}^k$, such that 
$(x_1,y_1) = (1,1)$, and for each $i \in \{2, \dots, k\}$ there exists 
$j < i$ such that $(x_i,y_i)$ is equal to one of the nearest neighbours 
of $(x_j,y_j)$, i.e. $(x_i,y_i) \in \cros_{x_j,y_j}$ and 
$(x_i,y_i) \ne (x_j,y_j)$.  In particular, the final coverage of the lattice 
can be computed as
\begin{equation}
N^\infty \equiv \lim_{t \to \infty} N(t) = \lim_{x \to 1} \Omega(x)
\label{asympcov}
\end{equation}
and other quantities of interest can be obtained by
(\ref{obvrelations}), (\ref{obvrelations2}) and (\ref{obvrelations3}).
Thus, the adsorption algorithm has been transformed to the problem
of finding the numbers of sequences in the sets $\setH_k$, $k=1,2, \dots.$
Once, we have the numbers $|\setH_k|$ we can write $\Omega(x)$ for
any $\alpha$ and compute $N(t)$, $N_p(t)$, $N_c(t)$ and $A(t)$ 
by (\ref{Nomegafunomega}), (\ref{obvrelations}) and (\ref{obvrelations2}), 
provided that the series in $\Omega(x)$ is convergent. It was pointed out 
in \cite{Fan:1991:UMS} that the convergence of series (\ref{Nomegafunomega}) 
is slow for $\alpha=1$ and for $x=1$. To overcome this difficulty,
we could use Shanks transformation or Pad\'e approximants as
in Section \ref{secalpha0}. This approach works in general and we will
use it in
Section \ref{secgeneralcase} where the general analysis of pRSA algorithm
is presented. Here, we present an alternative approach, rewriting
series (\ref{Nomegafunomega}) in different variables. Several possibilities 
were shown and motivated in \cite{Fan:1991:UMS}. 
Here, we write $\Omega(x)$ as
\begin{equation}
\Omega(x(z)) = \sum_{k=1}^\infty a_k z^k, 
\qquad
\mbox{where} \;\;\; x = \int_0^z \frac{3}{1 + 2 (1 - \xi)^3} \dxi.
\label{Omegainz}
\end{equation}
Let us define
\begin{equation}
h_k =  M^2 \, \frac{\big| \setH_k \big|}{k!}.
\label{defhk}
\end{equation}
To find $h_k$, one has to solve a finite combinatorial problem.
In this paper, we will make use of the first eight values of $h_k$. 
They can be computed as follow
$h_1 = 10 000;$ $h_2 = 20 000;$ $h_3 = 40 000;$
$h_4 \doteq 73 333;$ $h_5 \doteq 125 333;$ 
$h_6 \doteq 202 222;$ $h_7 \doteq 311 048;$ 
and $h_8 \doteq 459 452$.
To find coefficients $a_k$ in (\ref{Omegainz}), we substitute
$x \equiv x(z)$ in (\ref{Nomegafunomega}). We differentiate the resulting 
series term by term eight times and we evaluate each derivative
at $z=0$ to obtain:
\begin{eqnarray}
a_1 & = & h_1
\nonumber
\\
a_2 & = &  h_1 - \alpha h_2
\nonumber
\\
a_3 & = &  2 h_1/ 3 - 2 \alpha h_2 + \alpha^2 h_3
\nonumber
\\
a_4 & = &  h_1/ 6 - 7 \alpha h_2/3 + 3 \alpha^2 h_3 - \alpha^3 h_4
\label{coefak}
\\
a_5 
& = &  
-4 h_1/ 15 - 5 \alpha h_2/3 + 5 \alpha^2 h_3 - 4 \alpha^3 h_4 + \alpha^4 h_5
\nonumber
\\
a_6 
& = &  
-4 h_1/ 9 - 11 \alpha h_2/ 45 + 11 \alpha^2 h_3 /2 
- 26 \alpha^3 h_4 /3  + 5 \alpha^4 h_5 - \alpha^5 h_6
\nonumber
\\
a_7 
& = &  
- 20 h_1/ 63 + 6 \alpha h_2/ 5 + 53 \alpha^2 h_3 /15 
- 38 \alpha^3 h_4 /3  + 40 \alpha^4 h_5 /3 - 6 \alpha^5 h_6 + \alpha^6 h_7
\nonumber
\\
a_8 
& = &  
1.852 \, \alpha h_2 - 13  \alpha^2 h_3 /30 
- 63 \alpha^3 h_4 /5  + 145 \alpha^4 h_5 /6 - 19 \alpha^5 h_6 
+ 7 \alpha^6 h_7 - \alpha^7 h_8
\nonumber
\end{eqnarray}
Let $\overline{z}$ be a solution of equation 
$1 = 3 \int_0^z [1 + 2 (1 - \xi)^3 ]^{-1} \dxi$ (one can 
numerically estimate $\overline{z}$ as 0.569). Moreover,
let us denote 
\begin{equation}
\widetilde{\Omega} (z)
=
\sum_{k=1}^8 a_k z^k
\label{approxOmega}
\end{equation}
where $a_1$, \dots, $a_8$ are given by (\ref{coefak}). Then, 
using (\ref{Omegainz}) and (\ref{Nomegafunomega}), we can approximate
number of attached polymers as
\begin{equation}
N(t) \approx \widetilde{\Omega} \Big( z(t) \Big),
\; \;
\mbox{where $z(t)$ is given by} \;\; 
1-\exp \left[- \frac{t}{M^2} \right] 
= \int_0^{z(t)} \! \! \frac{3}{1 + 2 (1 - \xi)^3} \dxi.
\label{approxN}
\end{equation}
The results obtained by (\ref{approxN}) for $\alpha=1$ 
and $M=100$ are given in Figure \ref{figseries}(a).
To compute time evolution of $N(t)$, we chose an equidistant
mesh in $z$-variable in interval $[0,\overline{z}]$ and evaluated 
$\widetilde{\Omega}$ by (\ref{approxOmega}) at each 
$z$. The corresponding time $t$ was computed by (\ref{approxN}),
namely using the formula
$$ 
t 
= - M^2 \ln \left[ 
1 - \int_0^{z(t)} \! \! \frac{3}{1 + 2 (1 - \xi)^3} \dxi
\right].
$$
In Figure \ref{figseries}(a), we compare results obtained
by approximation (\ref{approxN}) and by stochastic
simulation of pRSA algorithm. We get a very
good agreement between the theoretically derived formula
and simulation. The asymptotic coverage can be approximated
as 
\begin{equation}
N^\infty \equiv \lim_{t \to \infty} N(t)
\approx 
\widetilde{\Omega}(\overline{z}),
\quad
N^\infty_c \approx \alpha \,  
\widetilde{\Omega}(\overline{z}),
\quad
\mbox{and}
\quad
N^\infty_p \approx (1 - \alpha) \, 
\widetilde{\Omega}(\overline{z}).
\label{NNcNpinfty}
\end{equation}
In Figure \ref{figseries}(b), we compare  approximations
of $N^\infty$, $N_p^\infty$ and $N_c^\infty$ computed by 
(\ref{NNcNpinfty}) with the results
of stochastic simulations. %
We see that approximations (\ref{NNcNpinfty}) 
provide good results for any $\alpha$. 
We can estimate relative error between approximation
$N_{app}^\infty$ and exact value $N^\infty$ as 
$(N_{app}^\infty-N^\infty)/N^\infty$. Using (\ref{NNcNpinfty}), 
we obtain $N_{app}^\infty - N^\infty \doteq 5$
and $(N_{app}^\infty-N^\infty)/N^\infty \doteq 0.14 \%$ 
for $\alpha = 1$. Here, $N^\infty$ can be computed by averaging 
over many realizations of the pRSA algorithm as 
$N^\infty \doteq 3641$ for $\alpha=1.$

In this section, we used transformation of variables 
(\ref{Omegainz}) to accelerate the convergence of series 
(\ref{Nomegafunomega}). This transformation was suggested 
in \cite{Fan:1991:UMS} for pRSA algorithm 
with $[\alpha,\beta] = [1,0]$, but our analysis shows that it can
give good results for any $\alpha$. 
The problem with this approach in general is determining
an appropriate change of variables. An easier, and more systematic, approach
is to use a Shanks transformation or Pad\'e approximants 
\cite{Shanks:1955:NTD,Wynn:1956:DCT,Weniger:1989:NST}
as we did in Section \ref{secalpha0}, and as we will do 
for the general analysis of the pRSA algorithm 
in Section \ref{secgeneralcase}.

\section{General analysis of pRSA algorithm}

\label{secgeneralcase}

\begin{figure}
\picturesAB{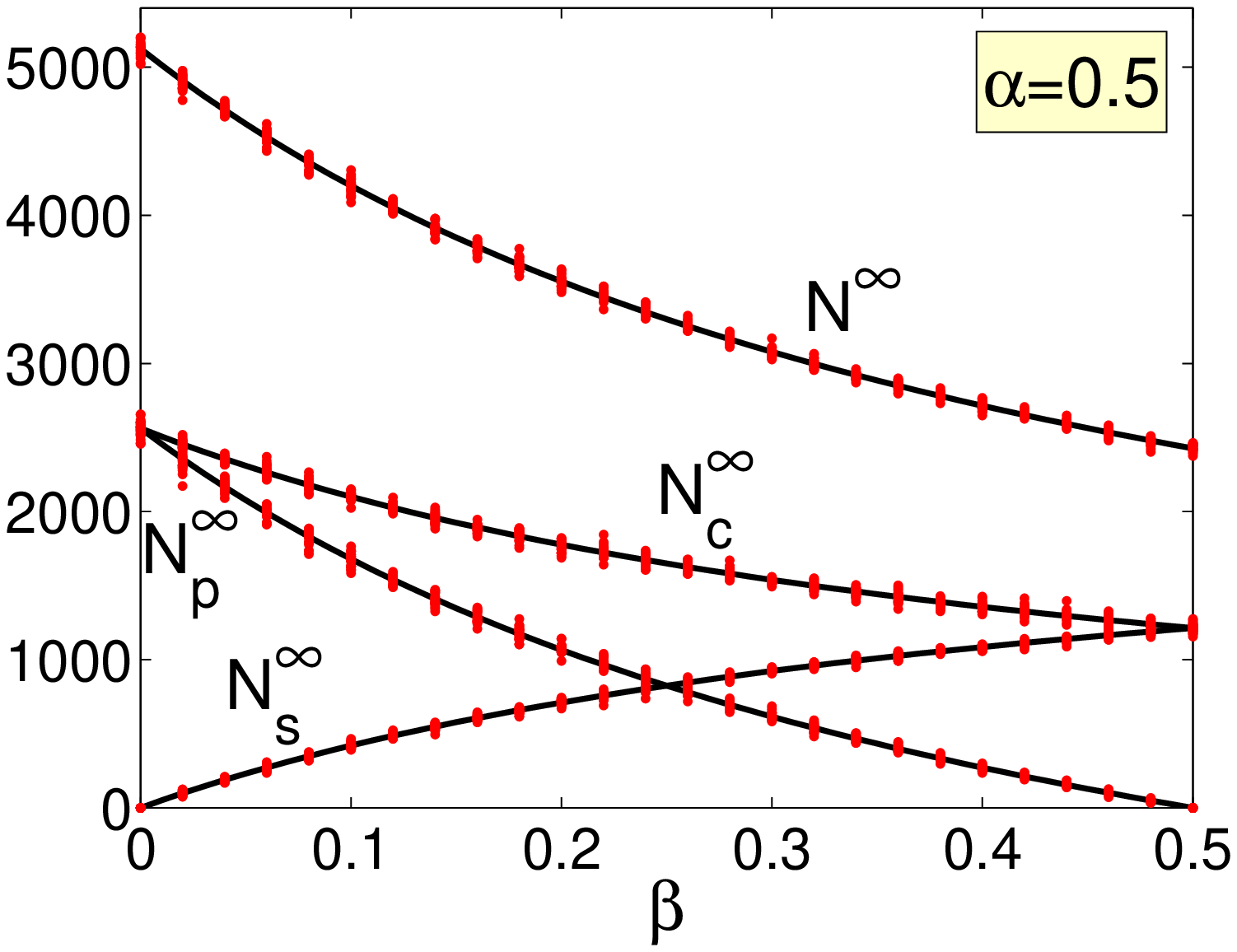}{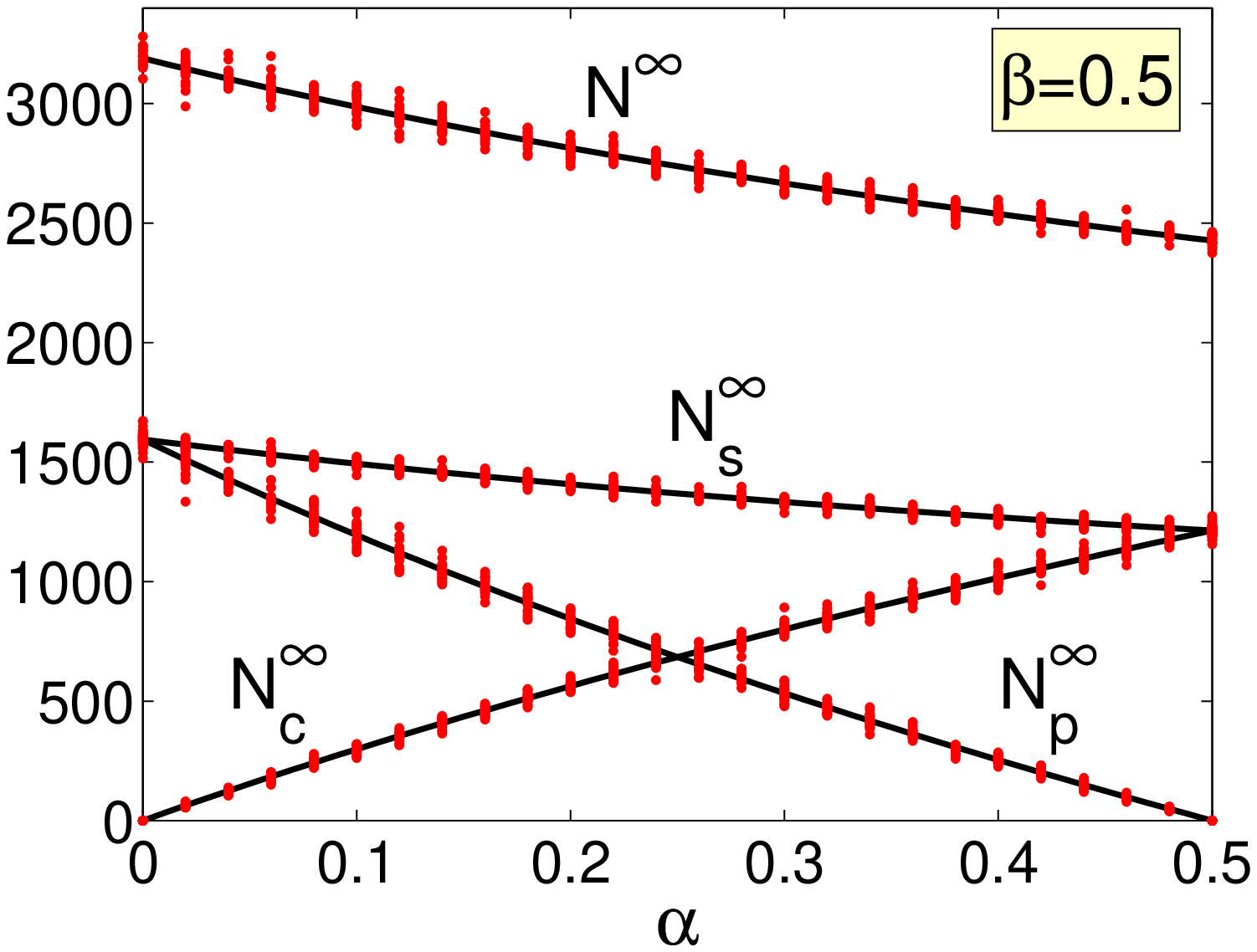}{2.07in}
\caption{
(a) {\it pRSA algorithm for $M=100$ and $\alpha=0.5$;} 
(b) {\it pRSA algorithm for $M=100$ and $\beta=0.5$.}
{\it In both cases, we present $N^\infty$, $N_p^\infty$, $N_c^\infty$ 
and $N_s^\infty$ as obtained by $(\ref{NNcNsNpinfty})$ (solid lines). 
We compare the approximate series expansion results with stochastic 
simulation of pRSA  algorithm (20 realizations, each realization plotted 
as a dot).}
}
\label{figseriesAlpha05Beta05}
\end{figure}

To evaluate (\ref{Ntseries}) for general $\alpha$ and $\beta$, we have 
to compute the quantities (\ref{combquan}) for $k=1,2,3,\dots$. Direct
evaluation of (\ref{combquan}) would require solving different 
combinatorial problems (weighted sums over all sequences in the set $\setP_k$) 
for different values of $\alpha$ and $\beta$. As in
Section \ref{secspecialcases}, we show that
a suitable reordering of terms can transform the set of combinatorial
problems to only one combinatorial problem which can be solved
independently of the values of $\alpha$ and $\beta$. Then the
dependence of the number of attached polymers and number of covered
binding sites can be easily studied. To do that, we first
use the Laplace transform to rewrite (\ref{Ntseries}) in terms of $\setG_k$. 
Here, as before $\setG_k$ is the set of all 
sequences of $k$ {\it distinct} points $\{(x_j,y_j)\}_{j=1}^k$, such that 
$(x_1,y_1) = (1,1)$, and for each $i \in \{2, \dots, k\}$ there exists 
$j < i$ such that $(x_i,y_i) \in \squ_{x_j,y_j}$ and 
$(x_i,y_i) \ne (x_j,y_j)$. 
Following a similar analysis to that in Section \ref{secalpha0}, 
we derive (compare with (\ref{Npsi}))
\begin{equation}
N(t) = \Phi \left( 1- \exp \left[ - \frac{t}{M^2} \right] \right)
\label{Nphi}
\end{equation}
where
\begin{equation}
\Phi(x) = M^2
\sum_{k=1}^\infty
\frac{\left( - 1 \right)^{k-1} x^k}{k!} 
\sum_{s \in \setG_k}
[\alpha + \beta]^{\xi(s)} 
\beta^{k - \xi(s) - 1},
\label{funphi}
\end{equation}
where, as before, $\xi(s)$ is the number of distinct points 
$(x_i,y_i) \in s,$  $(x_i,y_i) \ne (1,1)$, satisfying that there exists 
$j < i$ such that $(x_i,y_i) \in \cros_{x_j,y_j}$. Let $\phi_j^k$,
$j=1, \dots, k$, denote the number of sequences $s \in \setG_k$,
$k = 1, 2, \dots,$ satisfying $\xi(s) = k - j.$ The numbers $\phi_j^k$
for $k=1,2, \dots, 8,$ can be directly computed and they are given 
in Table \ref{tablephi}.
\begin{table}
\centerline{{\small
\begin{tabular}{|c||c|c|c|c|c|c|c|c|} \hline
 $\phi_j^k$  & $j=1$ & $j=2$ &$j=3$ &$j=4$ &$j=5$ &$j=6$ &$j=7$ &$j=8$ \\
\hline
\hline
$k=1$ & 1  & -- & -- & -- & -- & -- & -- & --\\
\hline
$k=2$ & 4  & 4 & -- & -- & -- & -- & -- & --\\
\hline
$k=3$ & 24 & 40 & 24 & -- & -- & -- & -- & --\\
\hline
$k=4$ & 176 & 424 & 424 & 176 & -- & -- & -- & --\\
\hline
$k=5$ &  1504  & 4800 & 6696 & 4776 & 1504 & -- & -- & --\\
\hline
$k=6$ & 14560  &   58368  &  104752  &  104280   &  57640   &  14560 & -- & --\\
\hline
$k=7$ & 156768 &  761024 & 1677680 & 2135920 & 1655336 &  745064 &  156768
& -- \\
\hline
$k=8$ &\!1852512\!&\!\!10603744\!\!&\!\!27833952\!\!&\!\!43206736%
\!\!&\!\!42818768\!\!&\!\!27137992\!\!&\!\!10289192\!\!&\!1852512\!\\ 
\hline
\end{tabular}}}
\caption{{\it Table of values of $\phi_j^k$ for $k=1,2, \dots, 8$,
$j=1, \dots, k$.}}
\label{tablephi}
\end{table}
Using the definition of $\phi_j^k$, formula (\ref{funphi}) can be rewritten to
\begin{equation}
\Phi(x) = M^2
\sum_{k=1}^\infty
\frac{\left( - 1 \right)^{k-1} x^k}{k!} 
\sum_{j=1}^k
\phi_j^k
[\alpha + \beta]^{k-j} 
\beta^{j - 1}.
\label{funphi2}
\end{equation}
Our task is to compute the sum of series (\ref{funphi2})
with reasonable precision, using only the first eight partial sums
$$
s_n (x) =  M^2
\sum_{k=1}^n
\frac{\left( - 1 \right)^{k-1} x^k}{k!} 
\sum_{j=1}^k
\phi_j^k
[\alpha + \beta]^{k-j} 
\beta^{j - 1},
\qquad
n = 1, 2, \dots, 8.
$$
To do that, we use Shanks transformation computed by Wynn's algorithm
(\ref{epsalgorithm}) and we approximate sum $\Phi(x)$
by term $\varepsilon_7^2(x)$, as in Section \ref{secalpha0}.
Thus we aproximate number of attached polymers as
\begin{equation}
N(t) \approx \varepsilon_7^2 \Big( x(t) \Big),
\; \;
\mbox{where $x(t)$ is given by} \;\; 
x(t) 
=
1-\exp \left[- \frac{t}{M^2} \right].
\label{approxNgen}
\end{equation}
The asymptotic coverage can be approximated as 
\begin{equation}
N^\infty \approx \varepsilon_7^2(1),
\quad
N^\infty_c \approx \alpha \,  \varepsilon_7^2(1),
\quad
N^\infty_s \approx \beta \,  \varepsilon_7^2(1)
\quad
\mbox{and}
\quad
N^\infty_p \approx (1 - \alpha - \beta) \, \varepsilon_7^2(1).
\label{NNcNsNpinfty}
\end{equation}
In Figure \ref{figseriesAlpha05Beta05}(a), we compare  approximations
of $N^\infty$, $N_p^\infty$, $N_c^\infty$ and $N_s^\infty$ computed by 
(\ref{NNcNsNpinfty}) with the results of stochastic simulations
for $\alpha=0.5$. The same plots for $\beta=0.5$ are given in
Figure \ref{figseriesAlpha05Beta05}(b). 
We see that approximations (\ref{NNcNsNpinfty}) provide excellent results.

\begin{figure}
\picturesAB{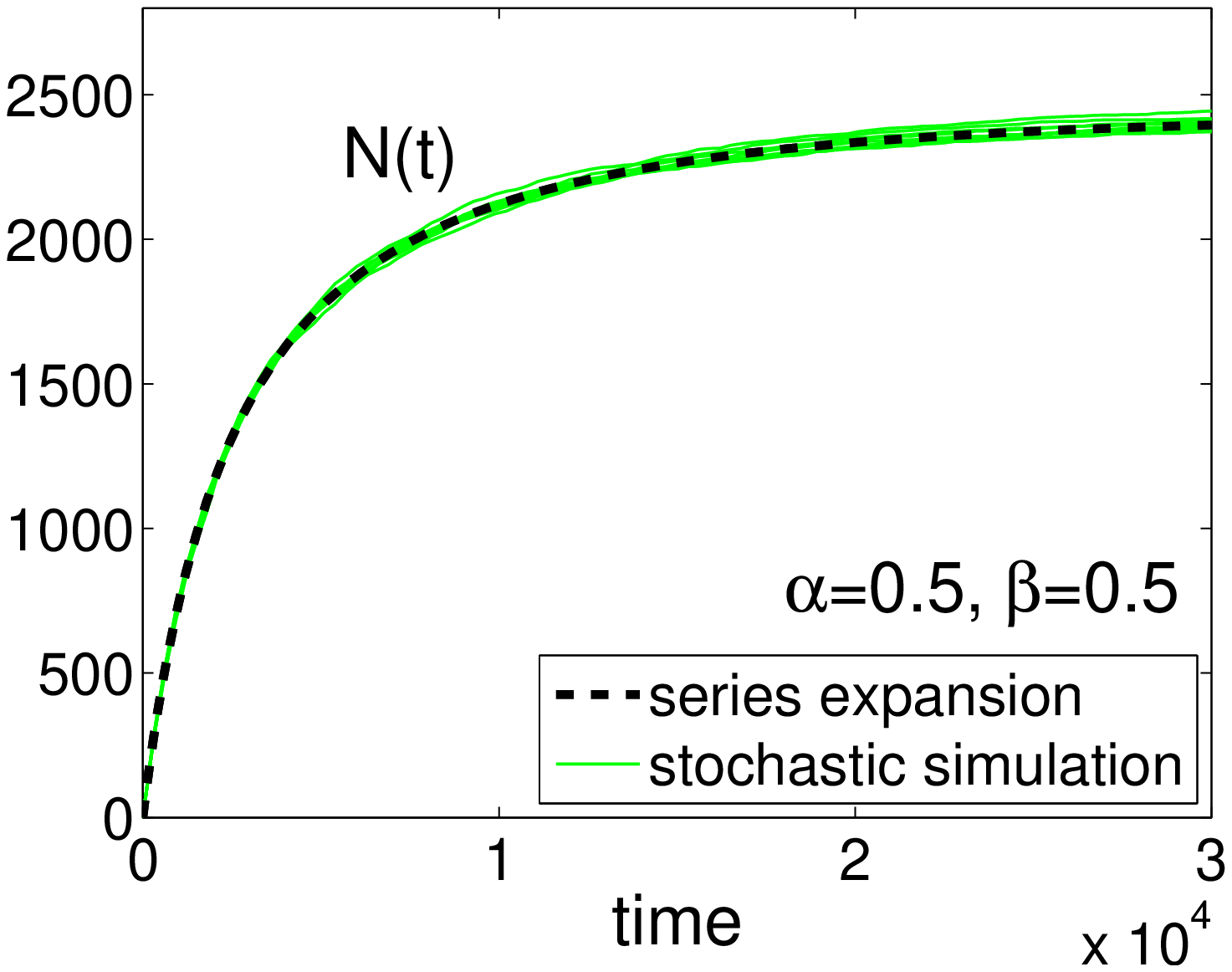}
{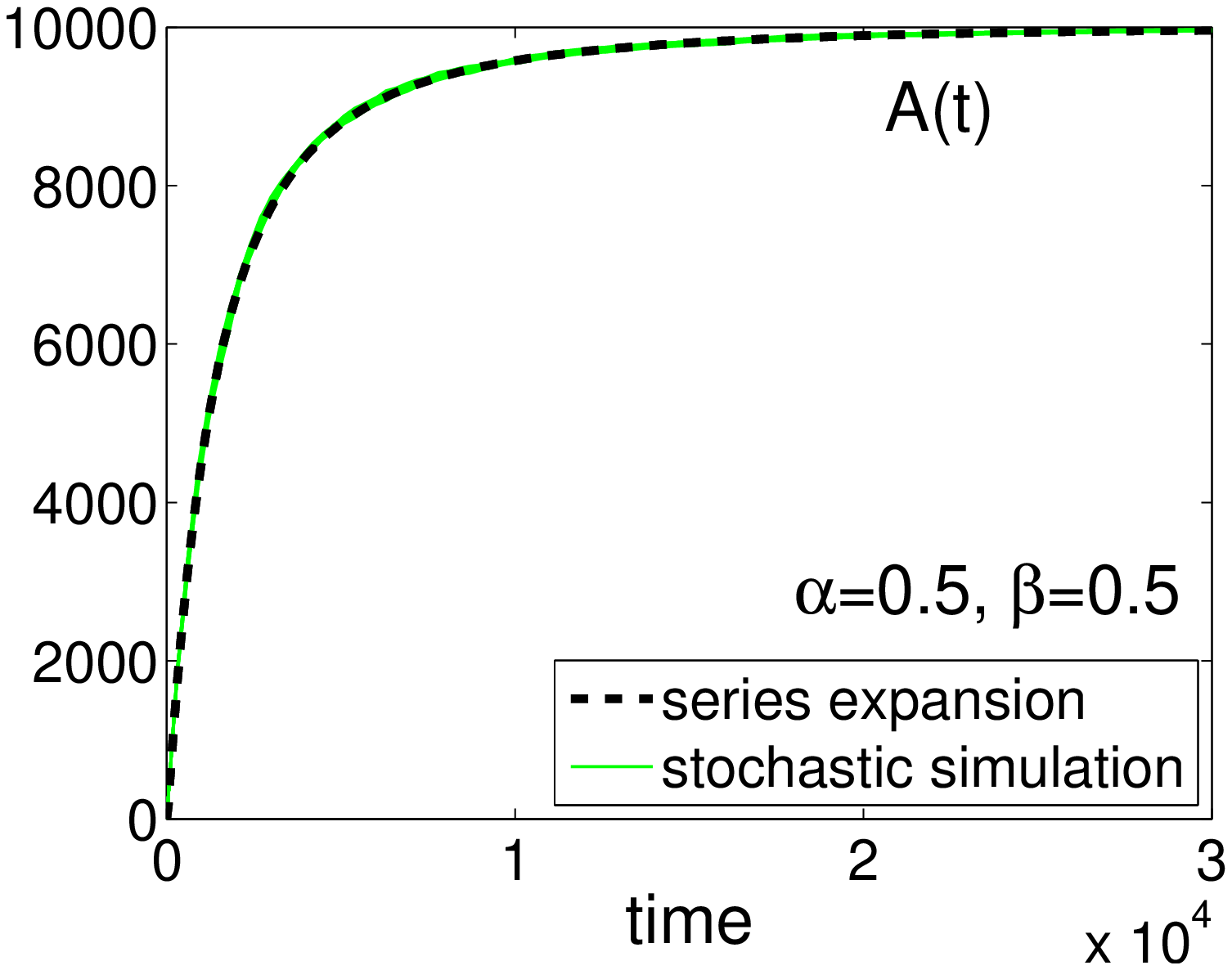}{2.07in}
\caption{{\it pRSA algorithm for $\alpha=0.5$, $\beta=0.5$ and $M=100$.} 
(a) {\it Time evolution of $N(t)$ given by
$(\ref{approxNgen})$ (dashed line). Ten realizations of stochastic 
simulation of pRSA algorithm are plotted as thin solid lines.}
(b) {\it Time evolution of $A(t)$ given by
$(\ref{approxNgen})$ and $(\ref{obvrelations2})$
(dashed line). Ten realizations of stochastic 
simulation of pRSA algorithm are plotted as thin solid lines.}
}
\label{figseriesalpha05beta05evol}
\end{figure}

\begin{figure}
\picturesAB{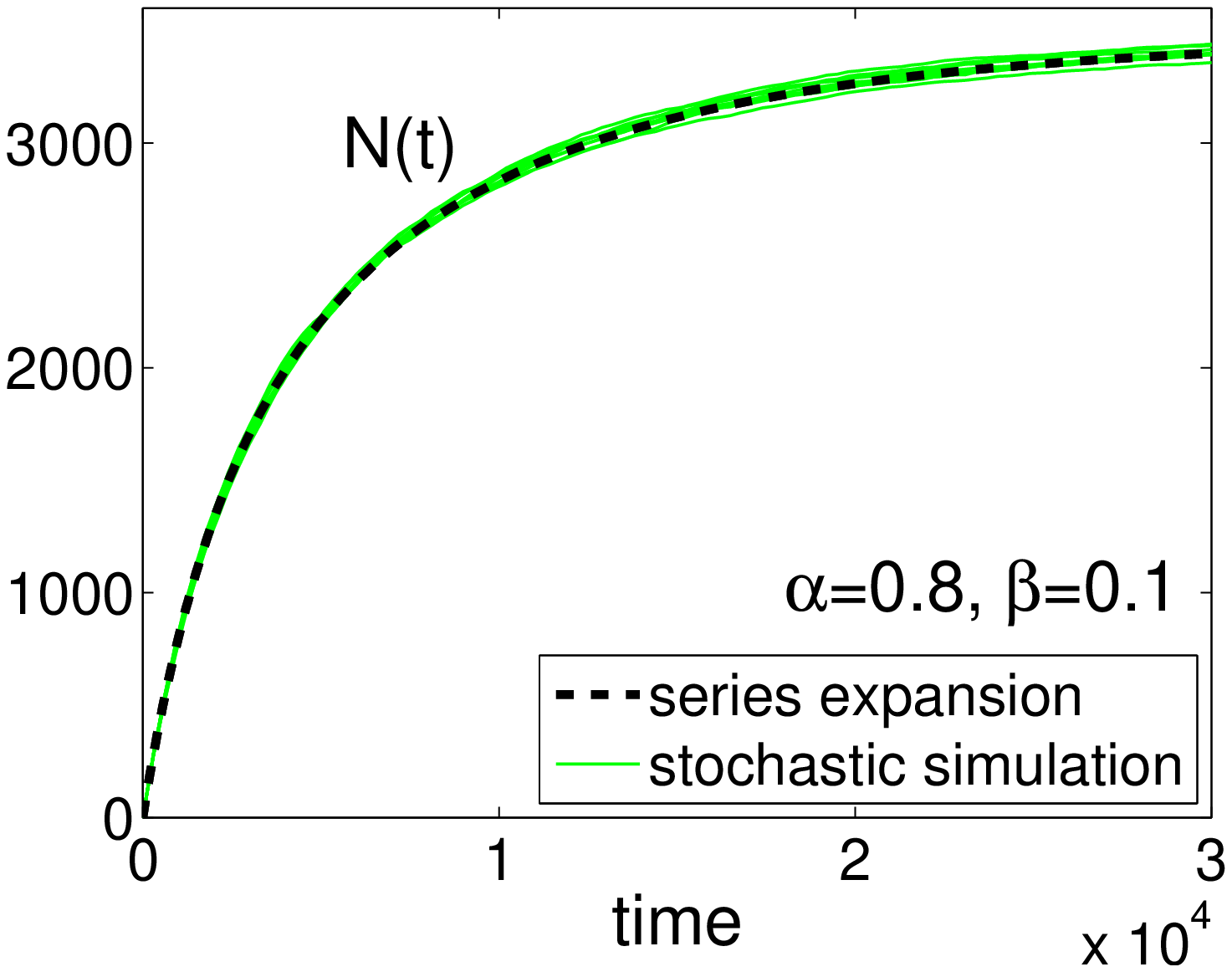}
{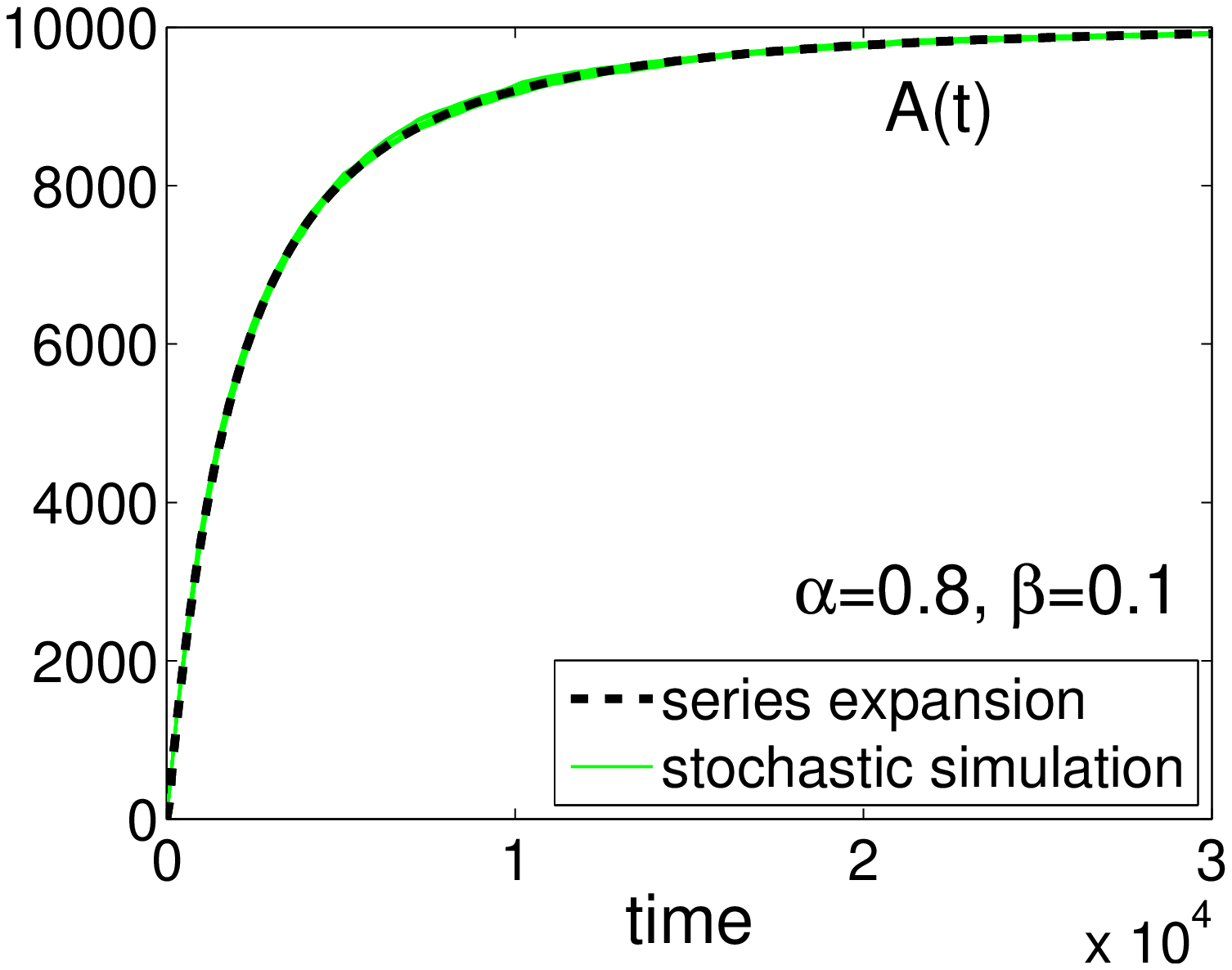}{2.07in}
\caption{{\it pRSA algorithm for $\alpha=0.8$, $\beta=0.1$ and $M=100$.} 
(a) {\it Time evolution of $N(t)$ given by
$(\ref{approxNgen})$ (dashed line). Ten realizations of stochastic 
simulation of pRSA algorithm are plotted as thin solid lines.}
(b) {\it Time evolution of $A(t)$ given by
$(\ref{approxNgen})$ and $(\ref{obvrelations2})$ (dashed line). 
Ten realizations of stochastic 
simulation of pRSA algorithm are plotted as thin solid lines.}
}
\label{figseriesalpha08beta01evol}
\end{figure}

The results obtained by (\ref{approxNgen}) for $\alpha=0.5$,
$\beta=0.5$ and $M=100$ are given in 
Figure \ref{figseriesalpha05beta05evol}.
To compute time evolution of $N(t)$, we chose an equidistant
mesh in $x$-variable in interval $(0,1)$ and evaluated 
$\varepsilon_7^2$ by (\ref{epsalgorithm}) at each 
$x$. Then the corresponding time $t$ was computed by (\ref{approxNgen}).
To compute $A(t)$ we used formula (\ref{obvrelations2}) where
the time derivative of $N(t)$ was approximate by the backward-in-time 
finite difference of $N(t)$.
In Figure \ref{figseriesalpha05beta05evol}, we compare results obtained
by approximation (\ref{approxNgen}) and by stochastic
simulation of pRSA algorithm. We see that we get a very
good agreement between the theoretically derived formula
and simulation. 
Finally, we present the time evolution of $N(t)$ and $A(t)$
for $\alpha=0.8$ and $\beta=0.1$ which is the situation shown
in the illustrative computation in Figure \ref{figal2surfplots0801}.
In Figure \ref{figseriesalpha08beta01evol}, we compare results obtained 
by (\ref{approxNgen}) with results obtained by stochastic simulation 
of pRSA algorithm. Again, we obtained an excellent
agreement between the series expansion results and the stochastic
simulation of the pRSA algorithm. 

\section{Discussion}

\label{secdiscussion}

In this paper we studied random sequential adsorption to 
the two-dimensional lattice. Our motivation
was chemisorption from polydisperse solution of polymers. 
We generalized the operator formalism of
\cite{Dickman:1991:RSA,Fan:1991:ACR}, derived series
expansion results and presented efficient methods 
to accelerate their convergence. In Section \ref{secalpha0},
we used classical methods for accelerating convergence of slowly 
converging series. In Section \ref{secbeta0}, we also 
presented results obtained by a more specialized transformation
of variables \cite{Fan:1991:UMS}. In both cases, the theoretical 
results compare well with the results of stochastic simulation of the
pRSA algorithm.

We assumed that the attached polymer can 
effectively shield a circle on the surface with radius $r < 2h$ 
where $h$ is the average distance between neighbouring binding sites.
We worked with the rectangular mesh of binding sites to enable the 
reformulation of the problem in terms of the RSA on the rectangular lattice. 
One should view this simplification as a reasonable approximation
of the problem where binding sites are more or less uniformly distributed 
on the surface. The restriction $r < 2h$ can be also relaxed
and the operator formalism could be generalized to the case of 
a mixture of longer polymers too. However, one should have in mind
that for larger $r$, the assumption that the ``wiggling tails"
of polymers can overlap has to be modified to take into account
the higher probability to find the polymer chain close 
to the binding site; see \cite{Doi:1986:TPD} for the general discussion
of the polymer dynamics.

Two-dimensional adsortption is more complicated to study because
there is no simple analogy of the exact approach which is 
available in one-dimension (see e.g. \cite{Evans:1993:RCS}
or the integro-differential evolution 
equation framework which was used in \cite{Erban:2006:DPI}). 
More precisely, one can formally write an evolution equation for the process 
(e.g. the master equation denoted (\ref{mastereq}) in this paper)
but it can be solved only by various approximation techniques 
\cite{Evans:1993:RCS}. For example, Nord et al \cite{Nord:1985:IMR} 
study adsorption of dimers or larger connected sites of objects 
to two-dimensional lattice. They write a master equation in hierarchic 
form for conditional probabilities that a conditioned configuration 
of mesh points is empty given that some neighbouring conditioning 
sites are empty. Using a series of hierarchic truncation schemes 
\cite{Vette:1974:KMD}, they were able to estimate dynamics and saturating 
coverage of the adsorption process. The operator formalism presented is 
a useful alternative to methods based on approximate evolution
equations.

The theoretical treatment of irreversible polymer adsorption is given
in \cite{Shaughnessy:2003:IPA}. They give a more detailed picture than
is studied in this paper, by studying the structure of the resulting
nonequilibrium layer in terms of the density profiles, and loop
and contact fraction distributions.
Adsorption of whole polymers to the surface, modelled as a self-avoiding
random walk, was done in \cite{Wang:1996:KJC} where the results
of Monte Carlo simulations are presented. It has been found that the coverage
to its jamming limit is described by a power law $t^{-\gamma}$ where
an exponent $\gamma$ depends on the chain length. In our case, we
modelled the adsorption of polymers as adsorption of disks
to the surface where the binding sites were arranged into the rectangular
lattice. In particular, the presented algorithm can be viewed as a
generalization of the classical lattice RSA models. Random sequential 
adsorption has been subject of the intensive research for the last sixty 
years. The reader can find more details about the RSA in review 
articles \cite{Evans:1993:RCS} and \cite{Talbot:2000:CPP}.

\providecommand{\bysame}{\leavevmode\hbox to3em{\hrulefill}\thinspace}
\providecommand{\MR}{\relax\ifhmode\unskip\space\fi MR }
\providecommand{\MRhref}[2]{%
  \href{http://www.ams.org/mathscinet-getitem?mr=#1}{#2}
}
\providecommand{\href}[2]{#2}

\end{document}